# Imaging Spectroscopic Ellipsometry of MoS$_2$

S. Funke[1*], B. Miller[2,3*], E. Parzinger[2,3*], P. Thiesen[1], A. W. Holleitner[2,3] and U. Wurstbauer[2,3]

[1]Accurion GmbH, Stresemannstr. 30, 37079 Göttingen, Germany

[2]Walter Schottky Institut and Physics Department, Technical University of Munich, 85748 Garching, Germany

[3]Nanosystems Initiative Munich (NIM), Schellingstr. 4, 80799 München, Germany

*contributed equally to the work

Micromechanically exfoliated mono- and multilayers of molybdenum disulfide (MoS$_2$) are investigated by spectroscopic imaging ellipsometry. In combination with knife edge illumination, MoS$_2$ flakes can be detected and classified on arbitrary flat and also transparent substrates with a lateral resolution down to 1 to 2 µm. The complex dielectric functions from mono- and trilayer MoS$_2$ are presented. They are extracted from a multilayer model to fit the measured ellipsometric angles employing an anisotropic and an isotropic fit approach. We find that the energies of the critical points of the optical constants can be treated to be independent of the utilized model, whereas the magnitude of the optical constants varies with the used model. The anisotropic model suggests a maximum absorbance for a MoS$_2$ sheet supported by sapphire of about 14 % for monolayer and of 10 % for trilayer MoS$_2$. Furthermore, the lateral homogeneity of the complex dielectric function for monolayer MoS$_2$ is investigated with a spatial resolution of 2 µm. Only minor fluctuations are observed. No evidence for strain, for a significant amount of disorder or lattice defects can be found in the wrinkle-free regions of the MoS$_2$ monolayer from complementary µ-Raman spectroscopy measurements. We assume that the minor lateral variation in the optical constants are caused by lateral modification in the van der Waals interaction presumably caused by the preparation using micromechanical exfoliation and viscoelastic stamping.

## 1. Introduction

Semiconducting transition metal dichalcogenides (TMDCs) such as MoS$_2$ belong to the emergent class of two-dimensional 'van-der-Waals' materials and are condensed in a hexagonal lattice structure with strong covalent in-plane bonds and weak van der Waals coupling between the individual planes [1-4]. The outstanding electronic [5,6], optical [7], catalytic [8,9], mechanical [10] , and spin- as well as valley properties [11] of semiconducting TMDCs cause significant interest for both fundamental research as well as device applications. It is characteristic to 2D materials that their properties are significantly modified by changing the number of layers. In the monolayer limit, the materials have often superior properties compared to their three-dimensional counterparts. For instance, the bandgap of MoS$_2$ undergoes a transition from indirect to direct semiconductor by thinning the material to the monolayer limit with a gap in the visible range [12] and two Raman active modes exhibit a sensitive thickness dependence [13]. Moreover, the photocatalytic stability is significantly increased in the monolayer limit compared to bi- and multilayers [14] making the material interesting for optoelectronic applications and solar energy harvesting. Absorption efficiency, optical transitions as well as excitonic properties are key to optoelectronic applications but also for fundamental studies. Thus, a detailed knowledge about the fundamental light-matter interaction is of great importance for basic research and application. The light-matter interaction is established by the complex dielectric function $\varepsilon(E) = \varepsilon_1(E) + \varepsilon_2(E)$ with real part $\varepsilon_1(E)$ and imaginary part $\varepsilon_2(E)$, respectively, or equivalently by the refractive index $n(E)$ and extinction parameter $k(E)$ all depending on the energy $E$. The dielectric function is a tensor entity. Since the properties of $MoS_2$ strongly depend on the number of layers, a layer-selective knowledge of the optical constants is demanding. Furthermore, it is inherent to 2D layered materials that the properties are highly anisotropic between in-plane (xy-plane) and perpendicular to the 2D crystals (z-direction). On the one hand a quantification of the full tensor of the complex dielectric function is crucial for an in-depth understanding of many optical phenomena. On the other hand, the lateral homogeneity of the optical constant on the micro and nanoscale is of practical relevance to micro- and nanoscale optoelectronic devices.

Thus far, the experimental determination of the complex dielectric function by spectroscopic ellipsometry of $MoS_2$ and other TMDCs has been reported for CVD grown samples [15–19]. In these studies, large areas are investigated averaging over several crystal domains due to the large excitation areas and presumably also over different number of layers depending on the growth method. Moreover, the in-plane component of the optical constants for different mechanically exfoliated TMDC monolayers have been extracted from reflectance measurements utilizing a Kramers-Kronig constraints analysis [20]. However, these do not provide access to the out-of plane components of the optical constants. So far, spectroscopic imaging (IE) ellipsometry with a micrometer resolution has been successfully applied to micromechanically exfoliated graphene [21, 22] and thermally reduced graphene oxide [23] for the identification on arbitrary substrate as well as to determine the optical constants with the required lateral resolution for detailed studies of the light matter interaction of two-dimensional materials.

In this paper, we report on spectroscopic imaging ellipsometry (SIE) at visible light frequencies and demonstrate that SIE is a powerful tool to detect and classify exfoliated $MoS_2$ mono- and few layer flakes on arbitrary flat substrate materials. It turns out that the extraction of the optical properties from spectroscopic ellipsometry measurements with high accuracy is difficult for $MoS_2$ supported by the standard $Si/SiO_2$ substrate material. The multilayer model needed to describe the measured ellipsometric spectra for such a layer sequence does not allow to clearly distinguish between the $MoS_2$-based signal and the contribution from the $Si/SiO_2$ substrate. We overcome this limitation by using transparent sapphire as substrate. For this purpose, an advanced technique to eliminate parasitic backside reflection of the polished rear surfaces of the transparent sapphire substrates is developed. The layer sequence consisting of sapphire, $MoS_2$ and air allows to clearly distinguish the contributions from substrate and $MoS_2$ flakes. This approach enables the determination of the optical constants with high accuracy and an excellent lateral resolution by SIE measurements down to 1-2 µm. We quantitatively compare the optical constants as well as the absorbance extracted from an anisotropic and an isotropic model to fit the spectroscopic ellipsometric data taken on mono- and trilayer $MoS_2$. Furthermore, we demonstrate that the optical constants close to the so-called A, B and C excitonic transitions feature a high lateral homogeneity over a few tens of microns, whereas the optical constants in an energy range between the B and C exciton exhibit some lateral variation. A detailed comparison with Raman measurements demonstrate that the minor variation in the optical constants in a narrow spectral range cannot be correlated to strain, disorder or lattice defects [24]. Only minor variations in in the charge carrier density are observable from Raman measurements, but the charge carrier landscape seems to be uncorrelated to the lateral fluctuations of the optical constants. We assume that the minor lateral variation in the optical constants are caused by lateral modification in the van der Waals interaction presumably caused by the preparation using micromechanical exfoliation and viscoelastic stamping. Since the van-der Waals coupling scales with *1/r$^6$*, with *r* the distance between the two interfaces, already minor fluctuations in the distance between the sapphire surface and the MoS$_2$ flake induced by the viscoelastic stamping to transfer the MoS$_2$ onto the substrate might be responsible for the observed small lateral fluctuations in the charge carrier density. Overall, the homogeneity over large regions and robustness of the optical properties are important for MoS$_2$-based hybrid structure and van-der Waals heterostructures.

## 2. Experimental Section

### 2.1. Sample Preparation and Detection

MoS$_2$ mono- and multilayer flakes are prepared by standard micromechanical exfoliation from naturally occurring bulk crystals (SPI Supplies) using an adhesive tape. The flakes are either directly transferred from the adhesive tape to the substrate surface or placed with the help of viscoelastic polydimethylsiloxan (PDMS) stamps allowing a lateral positioning with µm control (used for preparation of MoS$_2$ on sapphire substrate) [25]. Substrate materials are either *p*-type doped silicon (Si) with 285 nm silicon dioxide (SiO$_2$) on top or R-plane sapphire substrates. Ti/Au or silicon alignment marks facilitate the orientation on the substrates. The MoS$_2$ flakes are identified by optical microscopy using the interference contrast [26] that is high for MoS$_2$ on Si/SiO$_2$ substrate as shown in Fig. 1 (a) and rather poor but still present for MoS$_2$ on sapphire. Monochromatic reflectivity maps (*light energy of 2.54 eV)*) are accomplished in a back-scattering geometry using a 100x objective and by placing the sample on a x-y-z piezo stack with a closed-loop resolution of 1 nm (Physik Instrumente P-611.2 NanoCube XYZ-System) resulting in a combined lateral resolution better than 600 nm. The reflected light is recorded with a photodiode (Thorlabs PDA 100A – Si switchable gain detector).

### 2.2. Raman spectroscopy

Raman spectroscopy is carried out in back scattering geometry using the 2.54 eV emission line of a Kr-/Ar-ion gas laser as excitation source. The excitation power is kept below *P* < 1 mW to avoid damage of the flakes. The laser is focused with a 100x objective, and the sample is placed on the x-y-z piezo stack again achieving a lateral resolution better than 600 nm. The scattered light is dispersed by a single grating (1/1800mm) spectrometer (Acton SP-2560 from Princeton Instruments) and collected by a charge coupled device (CCD) camera. A steep long pass filter is used in front of the spectrometer to suppress the elastically scattered laser light. In the Raman measurements, we focus on the two prominent Brillouin zone center phonon modes $E^1_{2g}$ and $A_{1g}$ that are sensitive to the number of layers [13, 24: G, H, I] ]. The longitudinal optical phonon mode $E^1_{2g}$, an in-plane vibration, softens by increasing the number of layers, whereas the homopolar $A_{1g}$ mode, an out-of-plane oscillation, stiffens by increasing the number of layers. Consequently, the energy difference $\Delta E=|E(A_{1g})-E(E^1_{2g})|$ allows counting of the layers. Additionally, the $A_{1g}$ phonon mode is sensitive to doping and therefore to the environment [27, 28]. The µ-Raman characterization measurements are done at room temperature either in ambient conditions or in vacuum ($p < 10^{-5}$ mbar).

### 2.3. Spectroscopic imaging ellipsometry

Ellipsometric enhanced contrast micrographs (ECM) and spectroscopic imaging ellipsometry (SIE) measurement are performed with a spectroscopic imaging nulling ellipsometer EP4 (Accurion Gmbh, Göttingen) in ambient conditions at room temperature. For SIE in the visible range from 1.7 eV (729 nm) to 3.1 eV (400 nm), monochromatic light is provided by a laser-driven Xenon lamp and a grating based monochromator leading to a spectral width of the output line between ±0.024 eV and ±0.005 eV (± 1 nm). Spectroscopic data are obtained in the spectral range between 1.7eV (729 nm) and 3.1 eV (400 nm) with constant wavelength steps of 5 nm. The light path as well as the polarization state of the light in the applied geometry is sketched in Fig. 1(a). A large area of the sample is illuminated with a collimated light beam with a residual NA of about 0.018. The reflected light is collected with a lens system and recorded with a CCD detector. The lateral resolution of the set-up is only dependent on the lens system in the analyzation path and the pixel size of the CCD chip and constitutes between 1µm and 2µm. The geometry ensures that the deviations in the angel of incidence (AOI) and the angel under which the reflected light is recorded are negligible small while maintaining high lateral resolution [24: 1]

For SIE, the light is guided through a polarizer for linear polarization and then through a compensator to prepare elliptically polarized collimated light such that the light reflected from the sample is again linearly polarized. The reflected light is directed through a 20x or 50x objective and an analyzer to a CCD camera enabling a lateral resolution down to 1 µm. In a suitable coordinate system, the complex reflectance matrix is described by $\rho = r_p/r_s = \tan\Psi \cdot e^{i\Delta}$, with $r_{p(s)}$ the amplitude of the parallel (*p*) and orthogonal (*s*) components of the reflected light normalized to their initial amplitude (amplitude of incoming light) and the ellipsometric angles $\Psi$ and $\Delta$, respectively [29]. To obtain micrographs with ellipsometric enhanced contrast, the angle between polarizer and analyzer is kept fixed and the resulting live-view is recorded. In the SIE mode, the intensity of the reflected light is minimized by a 90° alignment of the analyzer with respect to the reflected light fulfilling the nulling condition [30, 31] for the selected region of interest (ROI). The ROI is a certain area on the sample under investigation and therefore, just a certain number of pixel on the CCD camera will be used to adjust for the nulling conditions. The ROI can be as small as the area imaged by one individual pixel on the CCD. Here, a ROI is defined such that selected pixels are binned to achieve a sufficient signal to noise ratio. The area of the sample including significant parts of the surrounding substrate has been divided into a grid of equally sized ROIs with a lateral size of about 2µm x 2µm. In this way, the $\Psi$- and $\Delta$- values are taken individually for different ROIs on the sample with a high lateral resolution given by the overall resolution of the optics in the reflected light beam. For SIE maps, the polarizer and analyzer-angles are determined by fulfilling the nulling condition for each ROI. The complex dielectric functions of $MoS_2$ are extracted from the measured ellipsometric angles $\Psi$ and $\Delta$ as an input of a Levenberg-Marquardt-fit based on the Berreman 4 x 4 matrix method [32] for multilayered films and Tauc-Lorentz as well as Lorentz approaches [24]. The multilayer stack consists of substrate with a finite surface roughness, $MoS_2$ and air [24]. As input data for this model we used either individual ROIs or, for larger homogenous areas on the sample, we averaged over few carefully selected ROIs. The described ellipsometry method is sensitive to the in-plane and out-of plane contribution of the dielectric response of the crystal. Therefore, the whole dielectric tensor is addressed in our ellipsometric investigations. This is a clear advantage of SIE compared to reflection measurements in backscattering geometry that provide only access to the in-plane contribution of the dielectric tensor ($\varepsilon_x, \varepsilon_y$) [20]. The taken SIE spectra are modeled using an isotropic ($\varepsilon_x = \varepsilon_y = \varepsilon_z$) and an anisotropic ($\varepsilon_x = \varepsilon_y \neq \varepsilon_z$) fit approach. In literature, ellipsometry data from TMDCs are typically treated with only an isotropic model [16–20]. Due to the small out-of plane dimension of 2D materials, the light-matter interaction is believed to be dominated by the in-plane component of the dielectric tensor. Nevertheless, the out-of plane component of the dielectric tensor is expected to be small but non-zero and additionally, it influences the result of the modelling for the in-plane components compared to a pure isotropic model. The thickness of a single layer $MoS_2$ is fixed to the theoretical value of 6.15 Å [33] in the multilayer model.

Generally, backside reflections become a problem when using thin transparent substrates. The incoming light is reflected at the backside of the substrate overlapping the reflected light from the top layer. The backside reflections cause low signals and they generate high measurement errors. Roughening, darkening, wedging or taping the backside of the substrate are commonly used techniques to avoid backside reflection in ellipsometry [34–38]. For measurements on transparent double sided polished sapphire substrate, we develop a special designed beam cutter arrangement to suppress parasitic signals due to backside reflection. The working principle of the beam cutter utilizing a knife edge is shown schematically in Fig. 1(b) together with a large view IE image of $MoS_2$ on sapphire.

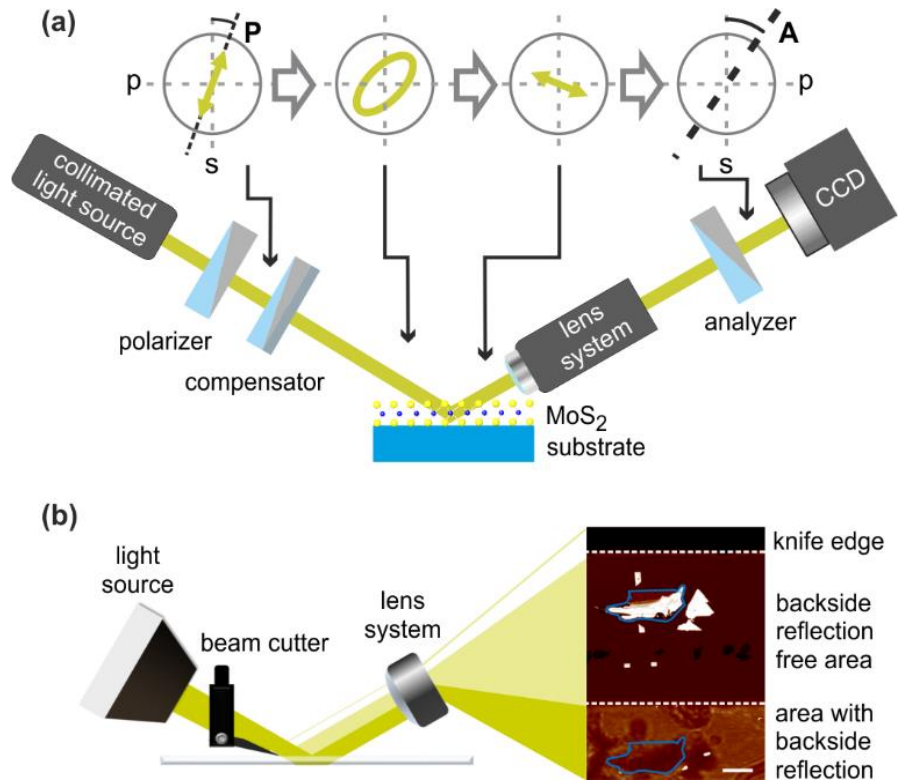

*Figure 1: Imaging ellipsometry setup. (a) Scheme of the imaging ellipsometry setup. The polarizer-compensator-sample-analyzer configuration for the reflected light is used. The sample is illuminated by collimated light. The additional lens system between sample and analyzer allows imaging with micrometer resolution. The lateral resolution is only limited by the optical system on the detector side, defined by the lens system and the CCD-camera. (b) The optional beam cutter alignment between light source and sample utilizing a knife edge suppresses effectively signal from reflections stemming from the back side of the transparent substrates. The picture on the right displays a large area view of the imaging ellipsometry contrast picture taken of a $MoS_2$ flake on double-sided polished sapphire using the beam cutter. In the dark region on top, the incoming light is blocked. In the middle area, reflections from the back side are selectively blocked with a sharp image of the $MoS_2$ flake and silicon alignment marker, while the lower part shows a blurred image of the $MoS_2$ flake that is a reflection from the flake on the back side of the substrate (scale bare denotes 50 µm).*

## 3. Results and Discussion

The ellipsometric enhanced contrast micrographs (ECM) and spectroscopic imaging ellipsometry (SIE) measurement on $MoS_2$ are carried out on exfoliated mono- and few-layer flakes supported either by $Si/SiO_2$ or transparent sapphire. The ECM and SIE data are compared with further imaging and spectroscopy measurements such as optical microscopy and monochromatic reflectivity maps as well as µ-Raman spectroscopy providing access to the number of layer, the strain and the lateral variation of the intrinsic doping level [27]. We will first compare the imaging methods and demonstrate that ECM is suitable to identify MoS$_2$ also on substrates providing poor contrast in optical microscopy. Next, SIE measurements will be introduced that are taken in the visible spectral range. The optical constants, particularly the complex dielectric functions $\varepsilon_1(E)$ and $\varepsilon_2(E)$ as well as the absorption coefficient $\alpha(E)$ are extracted from fits to the SIE data using an isotropic and anisotropic multilayer-model. The optical constants obtained from the two models are compared for mono- and trilayer MoS$_2$. Finally, the lateral homogeneity of the dielectric function is investigated for a large area monolayer MoS$_2$ on sapphire.

### 3.1. Imaging ellipsometry

Figure 2 displays optical microscopy (a), ECM (b) and a monochromatic reflectivity map (c) of a flake with mono-, bi-, few-layer and bulk parts of MoS$_2$ on Si/SiO$_2$ with an oxide thickness optimized for enhanced contrast in optical microscopy using white light illumination [26, 39, 40]. It is obvious that the optical contrast in Fig. 2(a) is good enough to unambiguously identify regions with a different number of layers. An ECM image of the identical flake is shown in Fig. 2(b). The image is taken with a 50x objective and the following set of parameters providing optimized contrast to distinguish between the different numbers of layers: angle of incidence of 50°, analyzer angle of -7.0°, polarizer angle of 34.6°, compensator angle of 45° and light energy of 2.82 eV. The color code reflects the real intensities captured by the CCD detector. Brighter regions correspond therefore to polarization changes of the reflected light that matches the analyzer angle better than compared to darker regions.

A comparison of Figs. 2(a), (b) and (c) demonstrates the capability of ECM, but also the one of monochromatic reflectivity in backscattering geometry to differentiate between mono-, bi-, and few-layer sheets. The difference between ECM and reflectivity maps in our experiments is that for the latter, the laser needs to be scanned over the region of interest (ROI) on the sample, whereas in ECM images the information is achieved with a single and thus very fast measurement. The number of layers are verified by Raman spectroscopy [Fig. 2(d)] from the energy difference between the two phonon modes $\Delta E=|E(A_{1g})-E(E^1_{2g})|$ utilizing the fact that the energy of the $E^1_{2g}$ mode decreased with increasing number of layers, whereas the $A_{1g}$ mode increases simultaneously [13]. In addition, the step height from mono- to the bilayer terrace has been determined by atomic force microscopy in tapping mode to be about 7Å, which is close to the reported interlayer distance in MoS$_2$ of 6.15 Å [33].

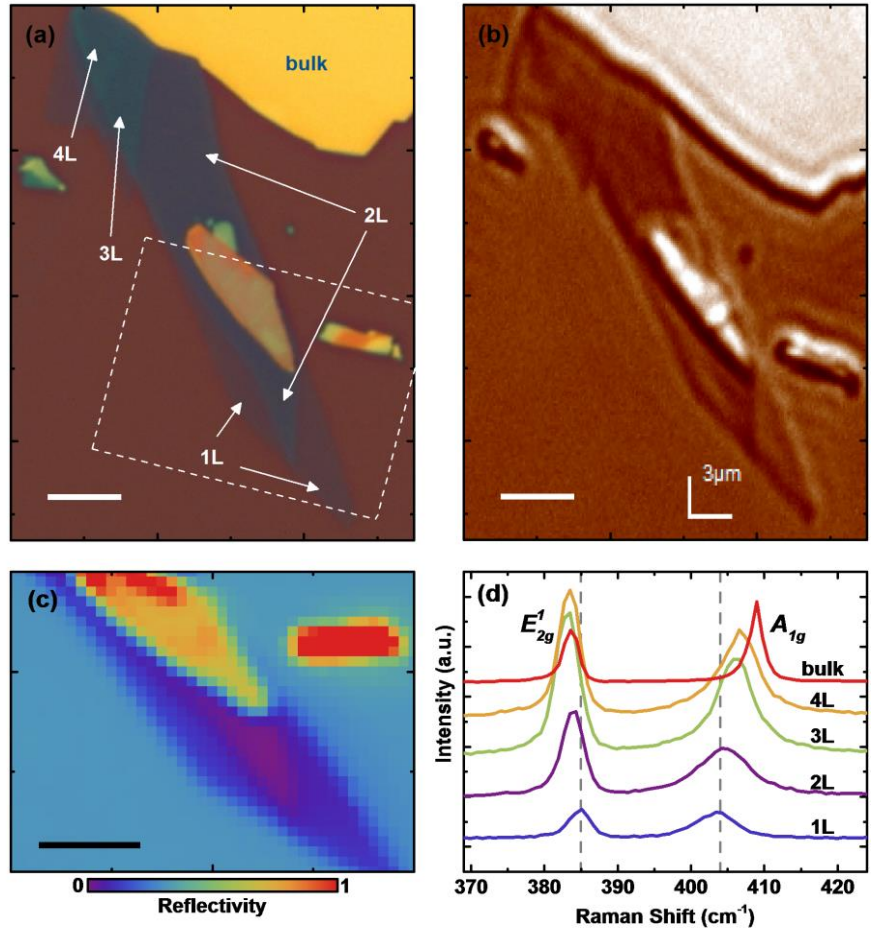

*Figure 2: Mono-and few layer MoS$_2$ flake on Si/SiO$_2$. (a) Optical microscopy image and (b) ellipsometric enhanced contrast micrograph (ECM) of the MoS$_2$ flake with mono-, bi-, tri- and multilayer regions. (c) Monochromatic reflectivity map (light energy of E = 2.54 eV) of the region marked with a box in (a). All scale bars denote 5 µm. (d) Raman spectra of 1L, 2L, 3L, 4L and bulk-like regions of the MoS$_2$ flake showing the $E^1_{2g}$ and $A_{1g}$ phonon modes. The difference of the mode energies verifies the number of layers [24 G,H].*

It is experimentally challenging to detect ultrathin MoS$_2$ flakes or other 2D crystals on arbitrary flat substrates and it is more complex to distinguish between mono-, bi-, and trilayer crystals by means of optical microscopy [26, 41–43]. We demonstrate the universality of ECM to characterize MoS$_2$ on transparent sapphire as supporting material. An ECM of a MoS$_2$ flake on transparent sapphire with an almost 50 µm long monolayer region as well as tri-, four- and multilayer terraces is displayed in Fig. 3. The bright slanted bar on the top of the flake is a Si-alignment marker (see open circle).

On such transparent substrates, the flake is barely visible by optical microscopy and it is nearly impossible to estimate the number of layers or to distinguish between the individual terraces. A magnified μ-Raman map of the boxed area in Fig. 3(a) is displayed in Fig. 3(b). The false color represents the layer-sensitive energy difference of the two prominent phonon modes $\Delta E=|E(A_{1g})-E(E^1_{2g})|$ [24: G,I]. The different terraces are unambiguously identified as mono- and tri-layer $MoS_2$. The fine wrinkle clearly seen in the ECM appears also in the Raman map at an energy belonging to a monolayer region. Regarding to the Raman measurements, it seems to be unlikely that the stripe is completely folded to an artificial tri-layer [44] and is therefore interpreted as wrinkled monolayer. The wrinkled region has a width of about 1 μm establishing the high lateral resolution of the imaging ellipsometry in ECM mode independent from the substrates similar to the findings for graphene [21].

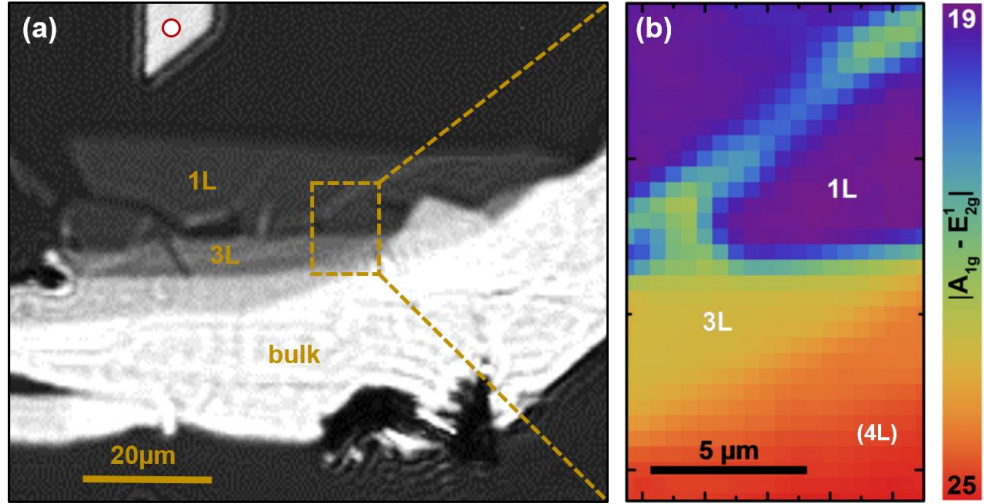

Figure 3: Mono- and few-layer $MoS_2$ flake on a transparent and double sided polished sapphire substrate. a) Ellipsometric enhanced contrast micrograph. (b) μ-Raman map of the region marked in (a). The energy difference $\Delta E=|E(A_{1g})-E(E^1_{2g})|$ between $A_{1g}$ and $E^1_{2g}$ phonon modes in units of wavenumbers ($cm^{-1}$) is color-coded identifying 1L, 3L and a 4L regions, but also a folded region in the 1L part. The folded region reveals a lateral resolution better than 1 μm in the graphs.

### 3.2. Spectroscopic imaging ellipsometry and layer dependent dielectric constants

Besides visualization of flakes, spectroscopic imaging ellipsometry (SIE) allows the determination of the optical properties, in particular the complex dielectric function $\varepsilon(E) = \varepsilon_1(E) + \varepsilon_2(E)$. In the energy range of the fundamental optical band gap $E_{gap} \approx 1.9$ eV, SIE maps exhibit a poor contrast between $Si/SiO_2$ and $MoS_2$. As a consequence, the optical constants for $MoS_2$ on $Si/SiO_2$ cannot be extracted with desired accuracy from a fit to the measured ellipsometric angles using the above described modeling procedure. For this reason, the presented SIE investigations are performed on $MoS_2$ on sapphire displayed in Fig. 3.

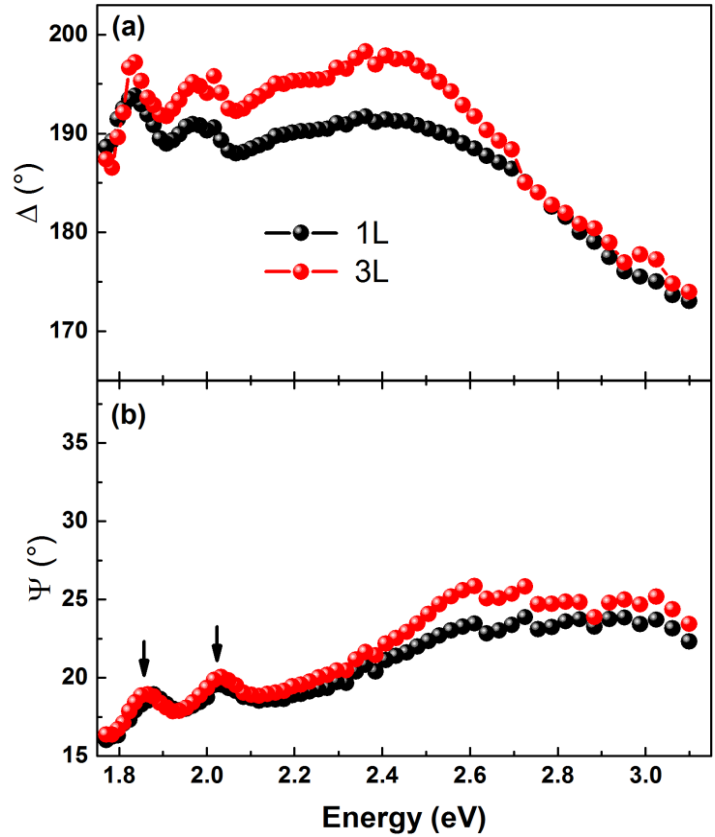

Figure 4: Ellipsometric spectra. $\Delta$-angle (a) and $\Psi$-angle (b) for mono- and trilayer MoS$_2$ on sapphire substrate in the energy range from 1.7 eV to 3.1 eV. The black arrows mark the spectral position of the "A" and "B" exciton in MoS$_2$.

In Figure 4(a,b) the determined ellipsometric angles $\Delta$ and $\Psi$ are plotted for mono- and trilayer MoS$_2$ on sapphire in the energy range from 1.75 eV to 3.1 eV. The $\Delta$- and $\Psi$-values shown for the monolayer are measured within the homogeneous regions of the MoS$_2$ flake displayed in Figures 3 and 7. It is evident from the experimentally observed ellipsometric angles that the critical points in the optical constants are observable at similar energies. Two rather sharp critical points are well resolved at an energy of around 1.85 eV and 2 eV [e.g. marked by arrows in Fig. 4(b)], respectively, and a broader peak can be determined at higher energies. The latter one seems to be more affected by the number of layers compared to the two critical points at lower energies. For a quantitative analysis and a comparison of the optical properties for different number of layers, it is necessary to fit the measured data to a suitable multilayer model as described in the experimental section 2.3.

Next, we focus on the comparison of the optical properties for mono- and trilayer MoS$_2$. The optical constants, i.e. the corresponding complex dielectric functions are extracted by fitting the ellipsometric spectra with an appropriate multilayer model assuming either an isotropic or an anisotropic dielectric tensor as described in details in the experimental section 2.3. The properties of van der Waals layered materials are highly anisotropic with respect to the in-plane (x,y-directions) orientation and out-of-plane (z-direction) orientation. For this reason, we apply both an anisotropic and an isotropic model to fit the ellipsometric $\Delta$- and $\Psi$- spectra that are taken under a finite angle of incidence of AOI = 50° and contrast the extracted optical constants. The ellipsometric angles of the sapphire substrate are measured simultaneously on a spot very close the MoS$_2$ flake. The modelled optical constants from the sapphire substrate show almost no dispersion in the investigated spectral range. The refractive index constitutes n = 1.7 in very good agreement with literature values. A finite surface roughness of the sapphire substrate is taken into account in the multilayer model. Five Lorentz profiles are used to fit the dielectric function with the isotropic approach. Similarly, five Lorentz profiles are used to fit the in-plane component of the dielectric function with the anisotropic approach. For the out-of-plane component of the dielectric function, a Tauc-Lorentz profile results in the best fit to the data. A Levenberg-Marquardt fit determines the best fitting parameters by minimizing the RMSE calculated as:

$$RMSE = \sqrt{\frac{1}{N_{exp}-N_{par}+1} * \sum_{j=1}^{M}\left\{\left[\frac{\Psi_{exp}(\lambda_j)-\Psi_{calc}(\lambda_j)}{\sigma\Psi(\lambda_j)}\right]^2 + \left[\frac{\Delta_{exp}(\lambda_j)-\Delta_{calc}(\lambda_j)}{\sigma\Delta(\lambda_j)}\right]^2\right\}},$$

with $N_{exp/par}$ describing the number of measured data points and fitted parameters respectively, $\sigma\Psi(\lambda_j)$ and $\sigma\Delta(\lambda_j)$ the measurement errors for $\Psi$ and $\Delta$, respectively, at a given wavelength. The RMSE for the best fit yields 5.6 for the isotropic and 2.1 for the anisotropic approach. The model describes the measured data sufficiently well for a RMSE in the order of 1 [45], whereas a model with RMSE >> 1 is insufficient to describe the experimental data. All fit-parameters and their results are provided in the supplementary information [24].

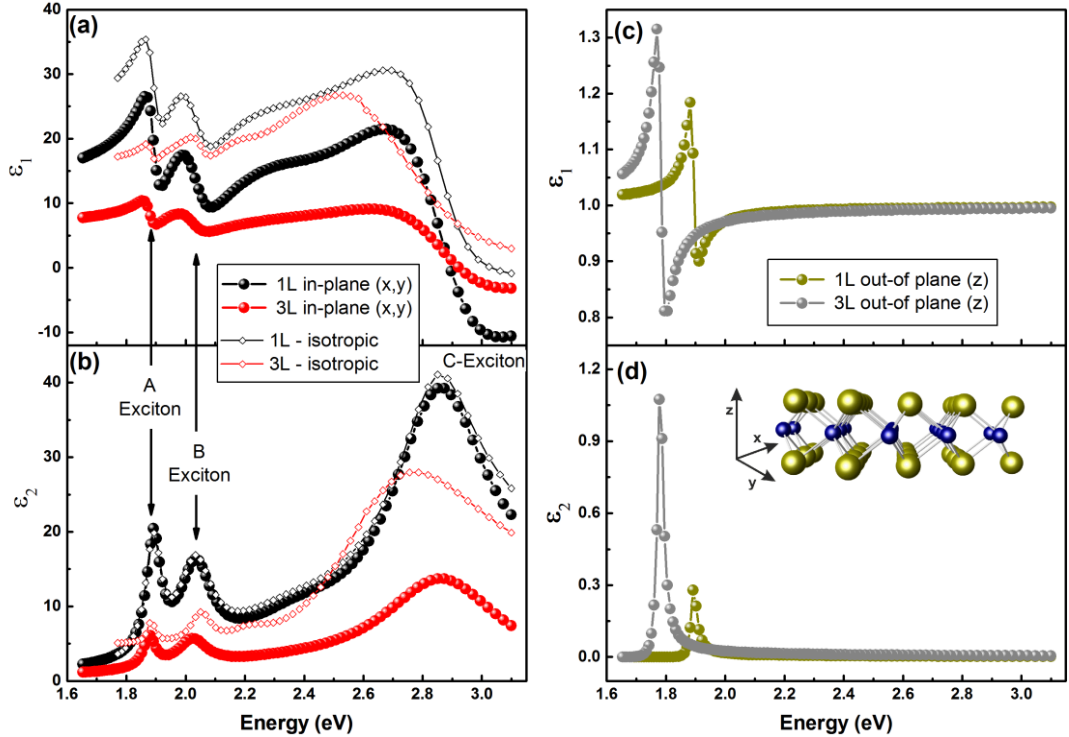

Figure 5: Complex dielectric function of MoS$_2$ extracted from fits to the ellipsometric data in Fig.4. (a) Real part of the in-plane component of the dielectric tensor ε$_1$(x,y) from an anisotropic model for monolayer MoS$_2$ (black solid spheres) and for a trilayer (red solid sphere). The dielectric function obtained from an isotropic model (open diamonds) is shown for comparison. (b) Imaginary part of the in-plane component of the dielectric tensor ε$_2$(x,y) as in (a) and the data from an isotropic model for comparison. (c): Real part ε$_1$(z) and (d): imaginary part ε$_2$(z) of the dielectric tensor in out-of plane direction from the anisotropic model to the ellipsometry measurements.

The real part ε$_1$ and imaginary part ε$_2$ of the dielectric tensor obtained from the isotropic and anisotropic fit approaches are summarized in Fig. 5. The real parts of the dielectric functions for the in-plane component ε$_1$(x,y) as well as the isotropic dielectric function are plotted in Fig. 5(a) and the imaginary part ε$_2$(x,y) in Fig. 5(b), respectively. There are three distinct critical points visible in the dielectric functions at around 1.9 eV, 2.05 eV and close to 3 eV. The critical points are assigned to the so-called A, B and C excitonic transitions [46]. The A and B excitons are transitions from the molybdenum (Mo) $d_z$-levels at the K(K') points in the Brillouin zone. At the K(K') point the valence bands are split by spin-orbit interaction resulting in the energy splitting of the A and B excitonic transitions [47] . The high joint density of states at higher energies causing a strong light matter interaction at the C-excitonic transition[48] , that is partially generated by band nesting, e.g. nearly parallel bands between M and Γ points in the Brillouin zone partially formed by hybridized molybdenum and sulfur *p*-orbitals [46, 47, 49] . We would like to mention that the Mo *d*-orbitals forming the Bloch bands at the K(K') points are well centered in the middle of the triple atomic layer forming the MoS$_2$ monolayer and is presumably only minor affected by changes in the environment such as van der Waals interaction and modification in the dielectric environment. In contrast, the sulfur (S) *p*-orbitals contributing to the band structure around Γ and M points [47, 48] are much more sensitive to changes in the environment. Since the van der Waals interaction scales with *1/r$^6$*, *r* is the distance between the two van der Waals interfaces, very little modification in the distance between the MoS$_2$ and supporting substrate might impact the bandstructure around the Γ and M points and, hence, the optical properties of the related energy range.

The energies of the critical points at the A, B and C excitonic transitions are well reproduced with both models for mono- as well as trilayer samples. The absolute magnitudes, however, differ significantly and they exhibit a clear layer-dependence as shown in Figs. 5(a,b). The magnitudes of the real as well as the imaginary part of the dielectric functions are found to be larger for the monolayer compared to the trilayer from both fit approaches. The comparison between the isotropic and anisotropic approach yields an almost identical imaginary part for the monolayer. The real part, however, is larger for the isotropic model in the whole investigated spectral range. The difference between the two fit approaches is better pronounced for the trilayer MoS$_2$. The magnitudes for the real and imaginary parts of the dielectric function are larger for the isotropic model compared to the anisotropic model. The energy of the critical point labelled C-exciton seems to be much more affected by the different models for the trilayer compared to the monolayer. The tendency that the differences between the models is enlarged for the trilayer MoS$_2$ compared to the monolayer might originate in the threefold thickness of the tri-layer, meaning that the out-of-plane component contributes more to the overall signal for the trilayer than for the monolayer.

For the monolayer, a quantitative comparison of the obtained dielectric function with values reported by Li *et al.* [20] for the in-plane component of the dielectric function yields better agreement with the data achieved from the anisotropic than from the isotropic model. There are no detailed data of the anisotropic optical constants for MoS$_2$ trilayers reported in literature.

The real part of the out-of-plane contribution of the dielectric function $\varepsilon_1(z)$ determined with the anisotropic model is plotted in Fig. 5(c) and the imaginary part $\varepsilon_2(z)$ in Fig. 5(d), respectively. There is only one distinct feature in the dielectric functions for each layer at photon energies slightly below the A-exciton feature. The magnitude of the dielectric function at this critical point below the A-exciton is much smaller compared to the values for the related in-plane contributions. The critical point is around 1.9 eV for the monolayer and therefore, it is close to the fundamental band gap. The critical point of the trilayer is at around 1.77 eV and hence, between the indirect gap (around 1.45 eV [12]) and the direct optical transition at the A exciton. The magnitude of the dielectric function at the critical point is almost three times larger for the trilayer compared to the monolayer. The real part of the out-of plane component of the dielectric function approaches a constant of about $\varepsilon_1(z) \approx 1$ and the imaginary part vanishes $\varepsilon_2(z) \approx 0$ for energies large than 2 eV likewise for mono- and trilayer MoS$_2$.

The knowledge on the dielectric function allows to calculate the absorbance $\alpha$ (Fig. 6). In particular, from the extinction coefficient *k* using the expression $\alpha = 4\pi k/\lambda$. Around 4% of the light irradiation perpendicular to the 2D plane is absorbed by a monolayer MoS$_2$ flake at the A and B exciton and almost 14% at the C exciton. The numbers agree well with current literature [12, 20, 50]. The absorbance is reduced in the whole visible range for the trilayer and constitutes around 2% at the A and B exciton and less than 10% at the C exciton [51]. This reduction of the absorbance between mono- and trilayer

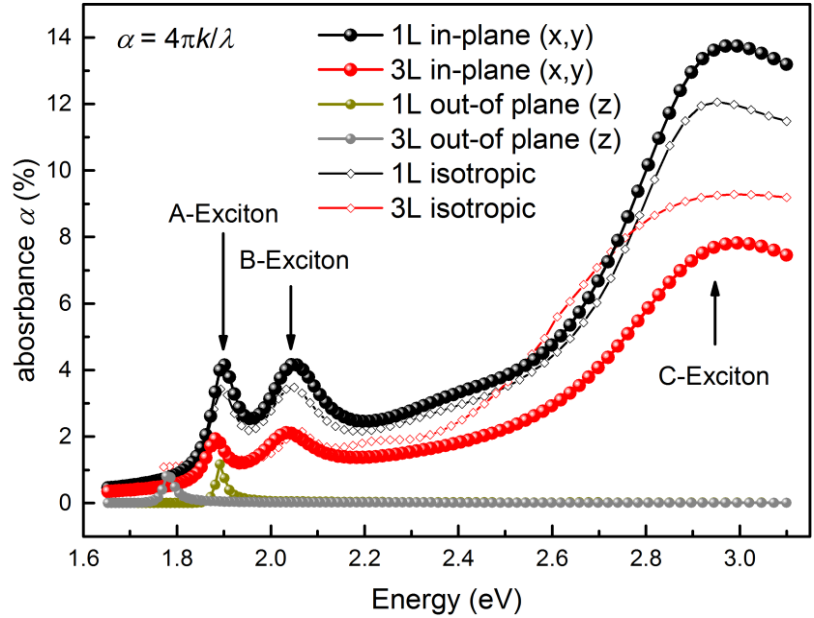

*Figure 6: Absorbance $\alpha$ for mono- and trilayer MoS$_2$ extracted from the optical constants, i.e. from the extinction coefficient in the visible range displaying high absorption efficiencies at A, B and C exciton transitions. The solid spheres are the results deduced from the in-plane component (xy-plane) from the anisotropic model (black mono-layer and red trilayer). The open diamonds represent the data of the isotropic model. The gray and green spheres represent the absorbance perpendicular (z-direction) to the 2D crystal determined from the anisotropic model (mono-layer green, tri-layer*

might be caused by the modified dielectric environment from pure air/sapphire as surrounding material in the monolayer case to an additional contribution from MoS$_2$ itself surrounding at least the inner MoS$_2$ layer of a trilayer. The tendency of the observed change in the absorbance is agreement with the reduced absorbance between monolayer MoS$_2$ and bulk material [20]. The absorbance of light irradiation parallel to the 2D material is 0.8% for the monolayer only at an energy of 1.9 eV and zero elsewhere. The absorbance for the tri-layer is about 1% for an energy of 1.77 eV and zero else.

The SIE measurements are consistent with a strong light-matter interaction of atomically thin monolayers. Particularly, it is stronger for the monolayer compared to the trilayer. Two different fit approaches have been employed to the measured ellipsometric angles taking into account either an anisotropic or an isotropic dielectric function. For the latter it is intrinsically assumed that the out-of-plane contribution to the dielectric function is negligible due to the height of the 2D crystal in the range of 1 nm. All critical points are well described at the appropriate energies for both models. The values for the dielectric function differ between the two models. Since MoS$_2$ belongs to the class of 2D materials that are highly anisotropic within the layer of crystal and perpendicular to it, it is meaningful to apply an anisotropic model to extract the dielectric tensor from spectroscopic ellipsometry investigations.

### 3.2. Lateral homogeneity of dielectric constant for monolayer MoS$_2$

The properties of MoS$_2$ can strongly depend on strain, doping, defects, interaction with the substrates and environment [27, 52–58]. We utilize the high lateral resolution of our imaging ellipsometry set-up to address the lateral homogeneity of the almost 50 µm x 10 µm large monolayer region of the MoS$_2$ flake on sapphire displayed in Fig. 3. To quantify the lateral homogeneity, we utilize the spatial distribution of the sum of the mean-square error (MSE) from fits to the ellipsometric $\Delta$ and $\Psi$ spectra using the isotropic fit approach as figure of merit. For the individual ROIs, the fit is performed in the whole measured spectral range, and the squared deviations between the measured $\Delta$- and $\psi$- values and the $\Delta$- and $\psi$-values determined from the fit at each energy are summed up. As shown in Fig. 7(a) and (b), we can separate the flake in two regions. The MSE of each pixel in region 'blue' (Fig. 7(a)) constitutes a MSE sum of error < 80 and in region 'red' (Fig. 7(b)) a MSE sum of 80 < error < 300. A histogram for the sum of MSE is displayed in the inset of Fig. 7(d). Representative $\Delta$- and $\psi$-spectra from both regions are plotted in Fig. 7(d). The overall difference is not very strong. At the critical points, the A, B and C excitonic transitions, almost no lateral variations can be identified. However, there is a minor deviation for $\Delta$ and $\psi$ at the A-excitonic transition (~1.9 eV). The largest difference can be found in the spectral range between B and C excitonic transitions. We do not find evidence that the pattern of the MSE coincide strongly with the morphology of the $MoS_2$ flake such as wrinkles or cracks.

To investigate the lateral homogeneity of the doping level, we display Raman spectra taken at a very low excitation power with special emphasize to the peak position and width of the $A_{1g}$ mode that is known to be very sensitive to the doping level [27, 59]. The Raman map shown in Fig. 7(d) displays the energy of the $A_{1g}$ mode in a false color code. For better orientation on the flake the Raman map is superimposed to a monochromatic reflectivity map of a larger area of the $MoS_2$ flake. With decreasing mode energy, the charge carrier density is increased (in the displayed region by roughly 1 order of magnitude [27]). There seems to be a charge carrier gradient from the inner part of the monolayer towards the monolayer edge with an accumulation of electrons at the free edge. Comparison of the MSE map with the density landscape determined from Raman measurements displayed in figure 7(c) and the monochromatic reflectivity map (figure 7(c)) does not provide any signatures which are directly correlated to each other. Consequently, the MSE variation seems to be independent of changes in the electron density. From further µ-Raman measurements, we do not find any evidence for a significant amount of strain, disorder or lattice defects that could be responsible for the lateral variation in the optical constants found in SIE. A detailed discussion of the Raman measurements is summarized in the supplemental material [24: G,H,I].

We surmise that the lateral variation in the MSE values dominating by the spectral range between B and C exciton is caused by some minor modifications in the van der Waals interaction between $MoS_2$ and the sapphire substrate. Those modifications are expected to be caused by minor fluctuations in the distance $r$ between $MoS_2$ and sapphire since the van der Waals interaction scales with $1/r^6$. This interpretation is supported by the fact that the A and B excitonic transitions - transitions at the K(K') points in the Brillouin zone, where the Bloch bands are predominantly formed by Mo d-orbitals that are well protected from the environment by the surrounding sulfur (S) atoms – are not affected by the lateral fluctuations of the optical constants. The minor fluctuations in the optical constants are only visible in the energy range between B and C exciton belong to transitions in $k$-space close to the M-point, where the sulfur $p$-orbitals contribute to the Bloch-bands. The sulfur $p$-orbitals and consequently also the electronic bands are highly sensitive to changes in the environment such as modification in van der Waals interaction. Due to the sensitivity of the sulfur $p$-orbitals to their environment, it is expected that also the optical properties in the energy range belonging to transitions in close vicinity to $\Gamma$ and M points are more affected by fluctuations e.g. in the van der Waals interaction compared to the A and B transitions at the K(K') points. Overall, the optical properties of the investigated monolayer part are characterized by a high lateral homogeneity. Within the resolution of complementary µ-Raman measurements, the landscape of the observed minor variations can neither be correlated with variations of the surface morphology, disorder, lattice defects in the $MoS_2$ crystal or with a variation of the electron density [24: G,H,I] .

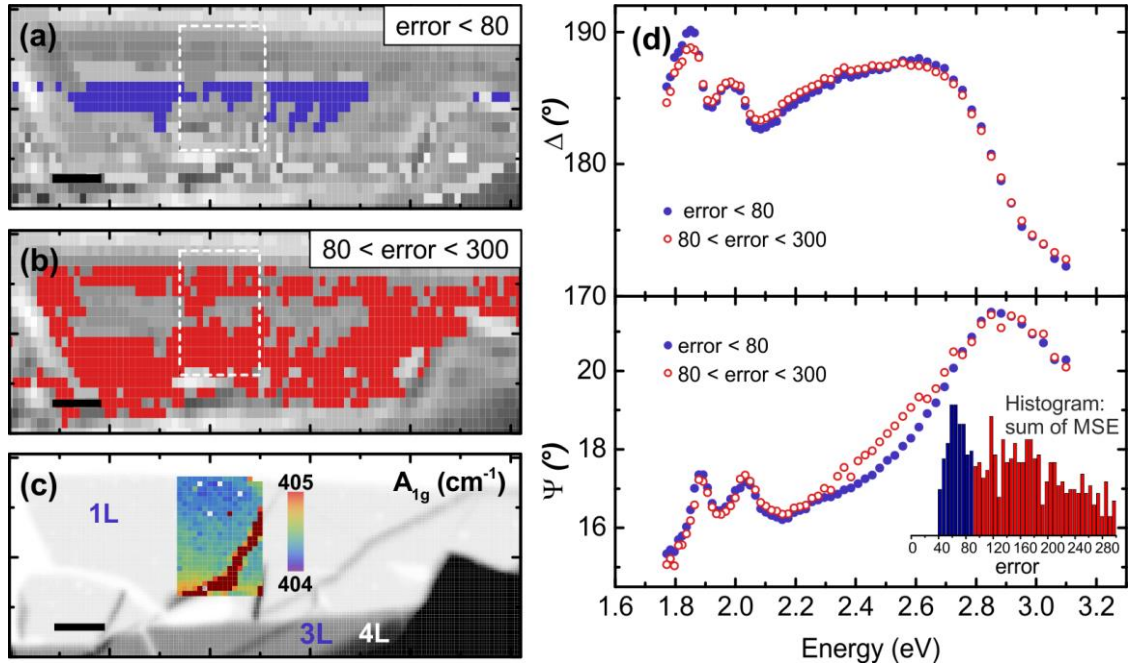

*Figure 7: Lateral homogeneity of the determined dielectric function for a large mono-layer region of MoS$_2$ on sapphire. (a) Blue region denotes a sum of mean-square errors MSE with a value below 80 determined from fits to the ellipsometry angles in the investigated spectral range. (b) Red region denotes a sum of mean-square errors 80 < error < 300. (c) Monochromatic reflectivity map (λ = 488 nm) and Raman map with the A$_{1g}$ mode energy color coded as a measure for the lateral variation in the charge carrier density. The scale bar denotes 10 μm. (d) Energy dependent Δ- and Ψ- values for representative positions with error < 80 (blue filled circles) and 8 0 < error < 300 (red open circles).*

## 4. Conclusion

In summary, we report on ellipsometric enhanced contrast micrographs (EMC) and spectroscopic imaging ellipsometry (SIE) of mechanically exfoliated MoS$_2$ flakes transferred to Si/SiO$_2$ and sapphire substrates. We show that EMC offers a good contrast for MoS$_2$ mono- and few-layers on both substrates. For substrates that are transparent and double sided polished an advanced technique is introduced to avoid backside reflections. The introduced beam cutter provides a reflection free area eligible to perform ellipsometric measurements on MoS$_2$, which is placed on arbitrary substrates. An outstanding lateral resolution of ~1 μm enables a layer-selective extraction of the optical constants from SIE measurements on high-quality exfoliated MoS$_2$ crystals. On the base of a Berreman 4 x 4 matrix method for multi-layered systems the SIE spectra are modeled. N-k-fix term, Lorentz- and Tauc-Lorentz-profiles are used to fit the ellipsometric Δ and Ψ spectra to extract the complex dielectric function.

Contrary to reflectance spectra, SIE is sensitive not only to the in-plane but also to the out-of-plane component of the dielectric function. We extract both components using an anisotropic model to fit the ellipsometric Δ and Ψ spectra. We compare the results with an isotropic modelling of the data. In the first approach, the out-of-plane component exhibits a significant contribution within a narrow energy range that changes with number of layers and it coincides with the indirect and direct band-gap for tri- and monolayer, respectively. The energies of the critical points, namely A, B and C excitonic transitions determined from both fit approaches are in excellent quantitative agreement. However, the magnitude of the real and imaginary part of the dielectric function determined by the two models are different. The deviation in the numbers is larger for the monolayer compared to trilayer part of the flake. The calculated absorbance is higher for the monolayer compared to the trilayer at all critical points.

The excellent spatial resolution of the imaging spectroscopic ellipsometry enables the study of lateral homogeneities of the dielectric constants. The homogeneity is characterized by the sum of MSE at each pixel. We find that the optical properties are robust and not significantly affected by wrinkle formations or lateral changes in the intrinsic charge carrier density. From further comparison with µ-Raman measurements we do not find evidence for a strain or a significant amount of disorder or lattice defects in $MoS_2$ crystal. The small lateral variations in the dielectric functions are observable in the spectral range between B and C exciton. This lateral variations are explained by laterally fluctuations in the van-der Waals interaction between $MoS_2$ flake and substrate affecting much more optical transition in k-space in the vicinity of $\Gamma$ and M points, where the band are also formed by sulfur *p*-orbitals, than transitions in k-space close to K(K') points related to the A and B excitonic transition, where the band-structure is dominated by Mo *d*-orbitals that are better protected from variations in the environment.

## Acknowledgements

We acknowledge support by Deutsche Forschungsgemeinschaft (DFG) via excellence cluster "Nanosystems Initiative Munich" (NIM) and through the TUM International Graduate School of Science and Engineering (IGSSE) and DFG projects WU 637/4-1, HO3324/9-1. This work is also supported by Federal Ministry for Economic Affairs and Energy on the basis of a decision by the German Bundestag and BaCaTeC.

# Supplementary Material: Imaging Spectroscopic Ellipsometry of MoS$_2$

S. Funke[1*], B. Miller[2,3*], E. Parzinger[2,3*], P. Thiesen[1], A. W. Holleitner[2,3] and U. Wurstbauer[2,3]

[1]Accurion GmbH, Stresemannstr. 30, 37079 Göttingen, Germany

[2]Walter Schottky Institut and Physics Department, Technical University of Munich, 85748 Garching, Germany

[3]Nanosystems Initiative Munich (NIM), Schellingstr. 4, 80799 München, Germany

*contributed equally to the work

## Setup and light path in imaging ellipsometry

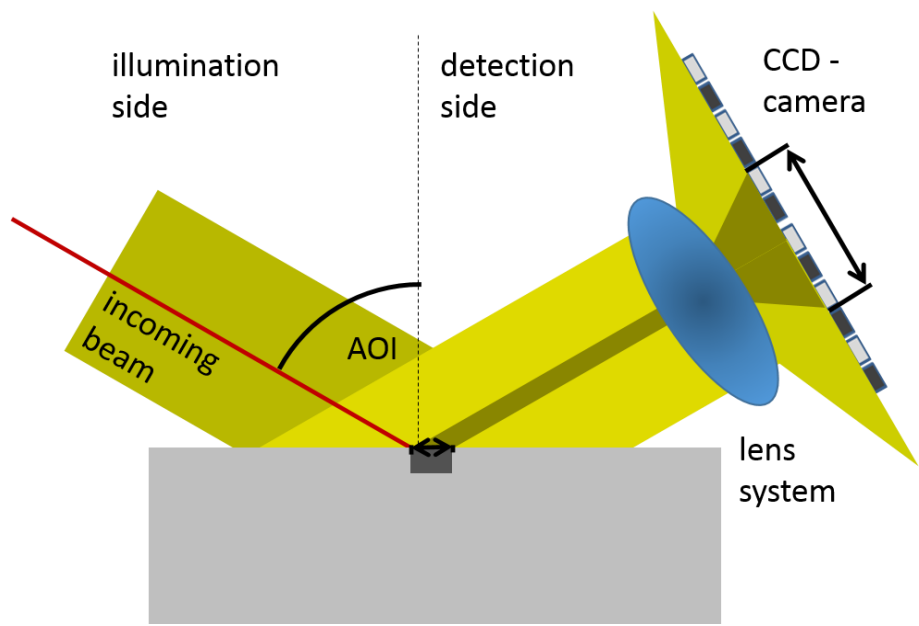

*Figure 1: Optical path in imaging ellipsometry. The collimated beam (NA ~0.018) illuminates a large area on the sample. The reflected light is guided through a lens system and displayed on the chip of a CCD detector. The resolution in imaging ellipsometry is determined by the lens system and the pixel size on the CCD chip and constitutes typically between 1 and 2 µm. The geometry ensures that the deviations in the angel of incidence and the angel under which the reflected light is recorded are negligible small while maintaining high lateral resolution.*

In figure 1 the light path in imaging ellipsometry is sketched with the focus on the optical elements providing the high lateral resolution. The sample is illuminated with a collimated light beam with a residual NA of 0.018. The reflected light is collected with a lens system and recorded with a CCD detector. The lateral resolution of the set-up is only dependent on the lens system in the analyzation path and the pixel size of the CCD chip and constitute in the current experiment between 1µm and 2µm. The geometry ensures that the deviations in the angel of incidence and the angel under which the reflected light is recorded are negligible small while maintaining high lateral resolution.

The optics to prepare and detect the elliptically polarized light and the change in polarization after reflection on the sample is introduced in detail in the manuscript figure 1 and the text in section 2.3.

# Model for fitting of the complex dielectric function

To fit the complex dielectric function of $MoS_2$, a layer structure as sketched in Figure 2 is defined. The exfoliated flake is placed on a sapphire substrate that exhibits some surface roughness. The roughness is taken into account by an additional layer with an effective thickness. The dispersion of the sapphire is extracted simultaneously from the ellipsometric measurement. The fit and resulting dispersion for sapphire and the layer accounting for the surface roughness are shown in Appendix C. The used dispersion formulas to extract the dielectric function for $MoS_2$ using the anisotropic and isotropic approach are introduced in Appendix A and B, respectively.

For the roughness of the sapphire substrate a thickness of 3.64 nm is determined from measurements and modelling of the bare substrate in close vicinity to the $MoS_2$ layer to avoid impact of spatial inhomogeneity. The determined roughness layer as well as the determined dispersion of the sapphire substrate material are used as input parameters in the fits to the data by the iso- and anisotropic approach, respectively. The thickness of the monolayer $MoS_2$ is set to 0.63 nm, whereas the thickness of the trilayer $MoS_2$ is set to 1.89 nm. The declared root mean square error (RMSE) in the paper (section 3.2) is taken as figure of merit to optimize the parameters. For the monolayer, the RMSE has improved from 5.570 to 2.148 by using the isotropic compared to the anisotropic fit approach. Similarly, the RMSE has improved from 10.099 to 3.084 by application of the anisotropic approach compared to the isotropic approach.

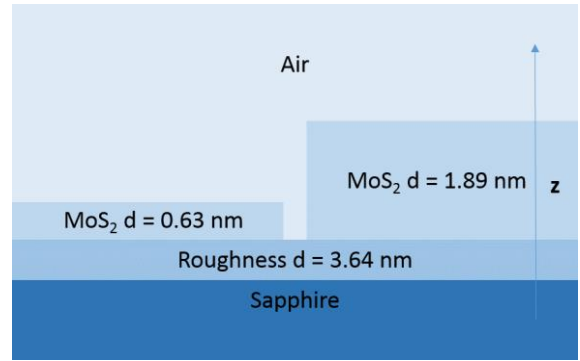

Figure 2: schematic figure of the MoS2 sample as used for the optical modelling

# Appendix A: Fit parameters and correlation matrix for anisotropic approach

We assume an anisotropic dispersion function for the $MoS_2$ in such a way that $\varepsilon_x = \varepsilon_y \neq \varepsilon_z$. The in-plane components ($\varepsilon_{x,y}$) of the complex dielectric function of $MoS_2$ is described by 5 Lorentz-oscillators and can be calculated by:

$$\varepsilon_{x,y}(E) = 1 + \sum_{i}^{n=5} \frac{s_i * f_i}{f_i^2 - E^2 - i*d_i*E},$$ where f describes the frequency of the oscillator in units of electron Volts (eV), $s$ is the strength of the oscillator, $d$ the damping and $E$ the photon energy in units of (eV). For the out-of-plane component, the imaginary part of the complex dielectric function is calculated by:

$$\varepsilon_{z,imag}(E) = \begin{cases} 0 & ; E \leq E_g \\ \frac{(E-E_g)^2}{E^2} * \frac{A*E0*\Gamma*E}{(E0^2 - E^2)^2 + \Gamma^2 * E^2} & ; E > E_g \end{cases},$$ where $E_g$ describes the bandgap energy, A

the strength of the oscillator at the energy $E0$ with the damping $\Gamma$. The model approach is implemented in the EP4Model software from Accurion following Ferlauto et. al.[1]. The fit results and its errors are shown in Table 1. Higher errors may occur if the correlation of the fitted parameters is high. The correlation describes the correlation between the parameters. It should be low for parameters of two different profiles, but may be higher for two parameters corresponding to the same oscillator term. The correlation of the fitted parameters is shown in Table 2. The color gradient is used to determine between low and high correlation.

The upper right triangular table denotes the correlation for the MoS$_2$ trilayer, whereas the lower left triangular table shows the correlation for the monolayer. The strong correlation of *s4* and *d4* for mono- and trilayer is given, because the fitted frequency of the oscillator is not in the measured range.

| Monolayer | | | | Trilayer | | | |
|---|---|---|---|---|---|---|---|
| | best fit | +/- | unit | | best fit | +/- | unit |
| f1 | 1.892 | 0.006 | eV | f1 | 1.878 | 0.001 | eV |
| s1 | 1.686 | 0.687 | eV$^2$ | s1 | 0.376 | 0.052 | eV$^2$ |
| d1 | 0 | 0.02 | eV | d1 | 2.024 | 0.002 | eV |
| f2 | 2.035 | 0.002 | eV | f2 | 0.049 | 0.006 | eV |
| s2 | 3.113 | 0.327 | eV$^2$ | s2 | 0.876 | 0.12 | eV$^2$ |
| d2 | 0 | 0.009 | eV | d2 | 0.121 | 0.012 | eV |
| f3 | 2.384 | 0 | eV | f3 | 2.374 | 0.076 | eV |
| d3 | 0.458 | 0.062 | eV | s3 | 2.257 | 2.490 | eV$^2$ |
| s3 | 4.853 | 1.065 | eV$^2$ | d3 | 0.81 | 0.448 | eV |
| f4 | 3.368 | 0 | eV | f4 | 6.001 | 4.069 | eV |
| s4 | 27.812 | 8.702 | eV$^2$ | S4 | 66.323 | 125.383 | eV$^2$ |
| d4 | 0.709 | 0.311 | eV | f5 | 2.873 | 0.004 | eV |
| f5 | 2.864 | 0.004 | eV | d4 | 3.006 | 9.295 | eV |
| s5 | 39.729 | 4.279 | eV$^2$ | s5 | 17.288 | 1.369 | eV$^2$ |
| d5 | 0.403 | 0.019 | eV | d5 | 0.479 | 0.02 | eV |
| A | 3.612 | 146.860 | eV | A | 7.699 | 15.868 | eV |
| E0 | 1.892 | 0.052 | eV | E0 | 1.779 | 0.003 | eV |
| Γ | 0.024 | 0.177 | eV | Gamma | 0.024 | 0.007 | eV |
| Eg | 1.810 | 2.366 | eV | Eg | 1.674 | 0.111 | eV |
| | | | | | | | |
| RMSE | 2.148 | | | RMSE | 3.084 | | |

*Table 1: Fitted parameters from the anisotropic approach. Results and its errors at final RMSE are shown for the monolayer (left) and the trilayer (right).*

| Trilayer \ Monolayer | A | E0 | Gamma | Eg | f1 | s1 | f2 | d1 | s2 | d2 | f3 | s3 | d3 | f4 | s4 | f5 | d4 | s5 | d5 |
|---|---|---|---|---|---|---|---|---|---|---|---|---|---|---|---|---|---|---|---|
| A | | 0.424 | 0.892 | 0.995 | 0.299 | 0.829 | 0.134 | 0.66 | 0.49 | 0.337 | 0.216 | 0.204 | 0.022 | 0.411 | 0.418 | 0.197 | 0.435 | 0.599 | 0.484 |
| E0 | 0.554 | | 0.045 | 0.339 | 0.227 | 0.314 | 0.007 | 0.3 | 0.044 | 0.06 | 0.005 | 0.02 | 0.058 | 0.114 | 0.112 | 0.001 | 0.094 | 0.083 | 0.055 |
| Gamma | 0.884 | 0.826 | | 0.925 | 0.191 | 0.708 | 0.141 | 0.535 | 0.531 | 0.38 | 0.222 | 0.201 | 0.047 | 0.467 | 0.473 | 0.204 | 0.485 | 0.653 | 0.523 |
| Eg | 0.972 | 0.725 | 0.963 | | 0.282 | 0.824 | 0.14 | 0.649 | 0.518 | 0.36 | 0.224 | 0.209 | 0.031 | 0.44 | 0.448 | 0.204 | 0.463 | 0.633 | 0.51 |
| f1 | 0.115 | 0.789 | 0.538 | 0.339 | | 0.248 | 0.119 | 0.204 | 0.223 | 0.21 | 0.034 | 0.048 | 0.015 | 0.107 | 0.108 | 0.049 | 0.112 | 0.155 | 0.122 |
| s1 | 0.763 | 0.91 | 0.964 | 0.891 | 0.718 | | 0.019 | 0.9 | 0.438 | 0.237 | 0.284 | 0.125 | 0.124 | 0.413 | 0.416 | 0.126 | 0.42 | 0.617 | 0.497 |
| d1 | 0.781 | 0.934 | 0.933 | 0.899 | 0.645 | 0.955 | | 0.067 | 0.075 | 0.064 | 0.1 | 0.207 | 0.166 | 0.218 | 0.224 | 0.145 | 0.234 | 0.219 | 0.181 |
| f2 | 0.046 | 0.545 | 0.301 | 0.187 | 0.55 | 0.385 | 0.441 | | 0.297 | 0.133 | 0.231 | 0.119 | 0.079 | 0.343 | 0.346 | 0.115 | 0.351 | 0.503 | 0.406 |
| s2 | 0.534 | 0.252 | 0.169 | 0.363 | 0.553 | 0.001 | 0.007 | 0.355 | | 0.911 | 0.198 | 0.379 | 0.596 | 0.033 | 0.022 | 0.231 | 0.011 | 0.289 | 0.229 |
| d2 | 0.367 | 0.279 | 0.068 | 0.223 | 0.463 | 0.05 | 0.082 | 0.328 | 0.921 | | 0.088 | 0.33 | 0.487 | 0.008 | 0.019 | 0.197 | 0.046 | 0.177 | 0.135 |
| f3 | 0.216 | 0.037 | 0.125 | 0.18 | 0.107 | 0.067 | 0.119 | 0.058 | 0.208 | 0.114 | | 0.542 | 0.609 | 0.227 | 0.219 | 0.436 | 0.223 | 0.434 | 0.551 |
| d3 | 0.273 | 0.276 | 0.312 | 0.3 | 0.158 | 0.298 | 0.302 | 0.058 | 0.241 | 0.304 | 0.404 | | 0.94 | 0.825 | 0.836 | 0.742 | 0.863 | 0.329 | 0.133 |
| s3 | 0.517 | 0.26 | 0.426 | 0.489 | 0.01 | 0.343 | 0.393 | 0.113 | 0.141 | 0.01 | 0.279 | 0.856 | | 0.609 | 0.622 | 0.698 | 0.66 | 0.102 | 0.015 |
| f4 | 0.275 | 0.054 | 0.1 | 0.201 | 0.251 | 0.009 | 0.078 | 0.141 | 0.383 | 0.302 | 0.316 | 0.002 | 0.148 | | 0.999 | 0.546 | 0.99 | 0.644 | 0.394 |
| s4 | 0.096 | 0.204 | 0.202 | 0.145 | 0.226 | 0.242 | 0.16 | 0.114 | 0.214 | 0.182 | 0.079 | 0.125 | 0.402 | 0.269 | | 0.576 | 0.995 | 0.668 | 0.424 |
| d4 | 0.011 | 0.163 | 0.132 | 0.064 | 0.227 | 0.185 | 0.098 | 0.121 | 0.257 | 0.21 | 0.053 | 0.19 | 0.491 | 0.281 | 0.99 | | 0.62 | 0.397 | 0.306 |
| f5 | 0.064 | 0.104 | 0.053 | 0.014 | 0.197 | 0.109 | 0.029 | 0.115 | 0.243 | 0.189 | 0.089 | 0.211 | 0.438 | 0.596 | 0.839 | 0.825 | | 0.682 | 0.447 |
| s5 | 0.113 | 0.152 | 0.058 | 0.039 | 0.283 | 0.138 | 0.042 | 0.135 | 0.412 | 0.34 | 0.291 | 0.019 | 0.35 | 0.523 | 0.935 | 0.928 | 0.89 | | 0.928 |
| d5 | 0.082 | 0.168 | 0.082 | 0.01 | 0.28 | 0.156 | 0.068 | 0.122 | 0.404 | 0.342 | 0.425 | 0.127 | 0.159 | 0.565 | 0.839 | 0.809 | 0.841 | 0.957 | |

Table 2: Correlation matrix from the anisotropic approach. Lower left triangular describes the correlation between parameters of the monolayer fit, whereas the upper right triangular displays the correlation between the parameters from fits to the data of the MoS$_2$ trilayer.

# Appendix B: Fit parameters and correlation matrix for isotropic approach

To fit the dispersion of Molybdenum-disulphide layer with the isotropic approach, Lorentz-profiles are used. The dispersion can be calculated by:

$$\varepsilon(E) = eps1 + \sum_{i}^{n=4} \frac{s_i * f_i}{f_i^2 - E^2 - i * d_i * E}$$

Table 3 summarizes the fit results for the mono- and the trilayer of MoS$_2$ using the isotropic approach. The RMSE, as described in the original paper, is better for the monolayer (5.570) as for the tri-layer (10.099). The correlation matrix is displayed in Table 4.

|      | Monolayer |       |       |      | Trilayer |       |       |
|------|-----------|-------|-------|------|----------|-------|-------|
|      | best fit  | +/-   | unit  |      | best fit | +/-   | unit  |
| f1   | 2.903     | 0.004 | eV    | f1   | 2.670    | 0.016 | eV    |
| s1   | 58.619    | 1.902 | eV²   | s1   | 34.722   | 2.927 | eV²   |
| d1   | 0.546     | 0.017 | eV    | d1   | 0.413    | 0.051 | eV    |
| f2   | 1.892     | 0.002 | eV    | f2   | 1.880    | 0.012 | eV    |
| s2   | 1.563     | 0.15  | eV²   | s2   | 0.435    | 0.264 | eV²   |
| d2   | 0.054     | 0.006 | eV    | d2   | 0.044    | 0.035 | eV    |
| f3   | 2.036     | 0.004 | eV    | f3   | 2.047    | 0.018 | eV    |
| s3   | 3.076     | 0.341 | eV²   | s3   | 1.061    | 1.144 | eV²   |
| d3   | 0.126     | 0.015 | eV    | d3   | 0.078    | 0.068 | eV    |
| f4   | 2.323     | 0.028 | eV    | f4   | 2.160    | 0.168 | eV    |
| s4   | 1.653     | 0.764 | eV²   | s4   | 0.481    | 1.955 | eV²   |
| d4   | 0.277     | 0.123 | eV    | d4   | 0.161    | 0.568 | eV    |
| eps1 | 10.426    | 0.3   |       | eps1 | 7.177    | 0.953 |       |
|      |           |       |       |      |          |       |       |
| RMSE | 5.570     |       |       | RMSE | 10.099   |       |       |

*Table 3: Parameters from isotropic fit approach and the related errors for the dispersion of MoS$_2$ mono- and trilayer..*

| Trilayer / Monolayer | f1 | s1 | d1 | f2 | s2 | d2 | f3 | s3 | d3 | f4 | s4 | d4 | eps1 |
|---|---|---|---|---|---|---|---|---|---|---|---|---|---|
| f1 | | 0.109 | 0.02 | 0.001 | 0.06 | 0.038 | 0.088 | 0.167 | 0.118 | 0.073 | 0.247 | 0.245 | 0.109 |
| s1 | 0.352 | | 0.784 | 0.009 | 0.013 | 0.019 | 0.164 | 0.273 | 0.183 | 0.121 | 0.485 | 0.456 | 0.569 |
| d1 | 0.212 | 0.871 | | 0.01 | 0.065 | 0.052 | 0.14 | 0.219 | 0.145 | 0.084 | 0.453 | 0.411 | 0.36 |
| f2 | 0.022 | 0.007 | 0.006 | | 0.022 | 0.022 | 0.004 | 0.044 | 0.064 | 0.018 | 0.007 | 0.008 | 0.024 |
| s2 | 0.029 | 0.105 | 0.101 | 0.174 | | 0.761 | 0.058 | 0.029 | 0.134 | 0.045 | 0.045 | 0.119 | 0.244 |
| d2 | 0.029 | 0.088 | 0.081 | 0.139 | 0.816 | | 0.044 | 0.057 | 0.127 | 0.013 | 0.011 | 0.065 | 0.178 |
| eps1 | 0.453 | 0.751 | 0.508 | 0.025 | 0.175 | 0.149 | | 0.592 | 0.512 | 0.582 | 0.539 | 0.42 | 0.062 |
| f3 | 0.046 | 0.084 | 0.074 | 0.134 | 0.293 | 0.249 | 0.023 | | 0.887 | 0.81 | 0.871 | 0.76 | 0.09 |
| s3 | 0.126 | 0.175 | 0.144 | 0.22 | 0.523 | 0.399 | 0.007 | 0.053 | | 0.664 | 0.701 | 0.555 | 0.055 |
| d3 | 0.094 | 0.123 | 0.105 | 0.204 | 0.524 | 0.373 | 0.004 | 0.012 | 0.89 | | 0.662 | 0.575 | 0.067 |
| f4 | 0.101 | 0.27 | 0.294 | 0.054 | 0.119 | 0.079 | 0.105 | 0.111 | 0.314 | 0.29 | | 0.932 | 0.204 |
| s4 | 0.195 | 0.61 | 0.633 | 0.07 | 0.213 | 0.154 | 0.266 | 0.214 | 0.579 | 0.463 | 0.153 | | 0.19 |
| d4 | 0.198 | 0.497 | 0.482 | 0.062 | 0.148 | 0.107 | 0.211 | 0.222 | 0.565 | 0.423 | 0.097 | 0.918 | |

Table 4: Correlation matrix for isotropic approach. Lower left triangular describes the correlation between parameters of monolayer fit, whereas the upper right triangular displays the correlation for the fitted parameters of the trilayer

# Appendix C: Fit of the sapphire substrate in close vicinity to MoS$_2$

A Cauchy term is used to fit the effective sapphire substrate. The measured data are obtained simultaneously with the data for the MoS$_2$. Table 5 shows the fitted parameters and its errors. For the roughness layer an effective-medium-approximation (EMA) Bruggeman is taken into account. With sapphire as host material and the guest material air with a fraction of 0.5, the EMA Bruggeman can be used to determine an effective roughness. Effective means, that the thickness of the EMA-layer may not correlate 1:1 to the real thickness. The fitted dispersion for the sapphire substrate as seen in Figure 3 is in good agreement with literature values of 1.7 – 1.8.

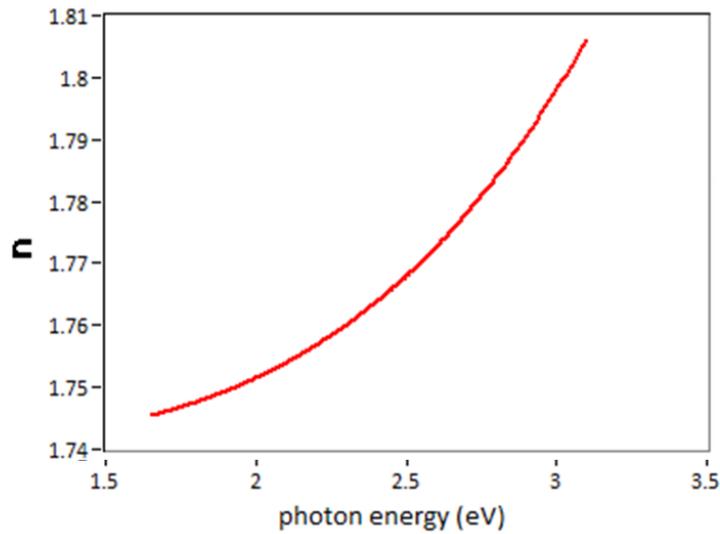

*Figure 3: Dispersion of the sapphire substrate from ellipsometry measurements in close vicinity to the MoS$_2$ flake. The determined sapphire dispersion is used as input for all models of the layer stack consisting MoS$_2$.*

| Sapphire | best fit | +/- | unit |
|---|---|---|---|
| thickness | 3.64 | 0.04 | nm |
| A_n | 1.740 | 0.001 | 1 |
| C_n | 1.69E+12 | 7.88E+10 | nm^4 |
| B_n | 0 | 2 | nm^2 |
|  |  |  |  |
| RMSE | 1.360 |  |  |

*Table 5: Fit parameters from the used Cauchy model to extract the dispersion of the sapphire substrate including a roughness layer at the surface.*

# Appendix D: Spectroscopic ellipsometry of a 4 layer $MoS_2$ part

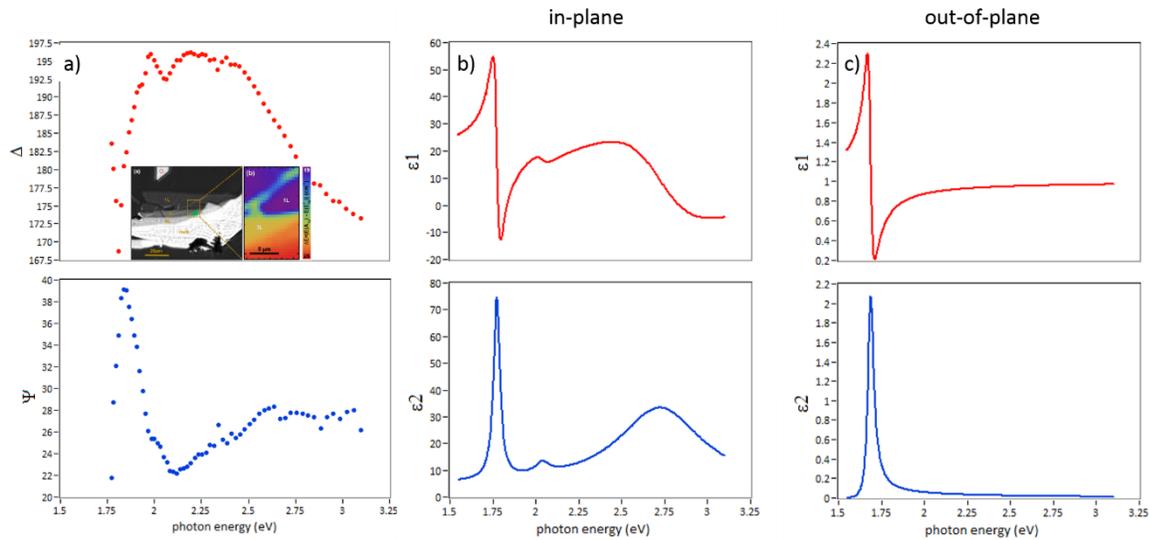

*Figure 4: (a) Measured ellipsometric angles for the small 4L part displayed in the inset. Due to step to a high multilayer region, the reflected light is impacted by parasitic reflection from the side facet of this step. The impact is most pronounced for small photon energies and reduced for larger photon energies. From fits to the measured data, the in-plane component (b) and the out-of-plane component (c) of the dielectric function for this 4L layer $MoS_2$ flakes are extracted from the measured data with the parasitic refelction from the side facet. The energies for the critical points in the dispersions seems to be reasonable, whereas the amplitude of the dielectric function are strongly affected by the parasitic signal from the site facet of the step and do not resemble the values for a pure 4L $MoS_2$ flake.*

In figure 4 (a) the ellipsometric angles Δ and Ψ values extracted from green area – a very small 4L region (<2μm) with a step to a thick bulk part adjacent to it - as shown in the inset. A magnified view of the imaging ellipsometry contrast picture and the related Raman map of the $MoS_2$ flake is displayed in figure 9. We would like to highlight that particularly for small photon energies the signal of the determined ellipsometric angles is significantly impacted by reflections from the side facet of the step to the adjacent balk part due to the finite angle of incidence and the limited lateral resolution of about 2μm. For completeness, the anisotropic fit approach was also performed for this small 4L region and the extracted in-plane and the out-of-plane components of the complex dielectric function are plotted in figures 4 (b) and (c), respectively. The obtained RMSE value of 19.353 is very high. All fitted parameters are shown in Table 6. The errors occur from high correlation and the additional reflection from the side facet of the step to the bulk part adjacent to the 4L region of the $MoS_2$ flake. The isotropic approach did not fit at all and is not shown. Below, we also show data of the fit to the model using only the ellipsometric angles Δ and Ψ as input in the anisotropic model for comparison from a window with larger photon energies (grey data points and lines traces in figure 7). Nevertheless, the extracted dielectric function of the 4L part remains imprecise and for small photon energies inaccurate. Despite the parasitic reflection signal from the step height in the measured ellipsometric data, the energies of the critical points at the A and B exciton close to the band gap but also for the higher energetic C-exciton are reasonable and follow the trend by going from mono- to tri- to four- and to few-layer $MoS_2$.

| Fit Results | | | |
|---|---|---|---|
| | best fit | +/- | unit |
| f1 XY | 2.738 | 0.063 | eV |
| s1 XY | 48.040 | 23.743 | eV$^2$ |
| d1 XY | 0.602 | 0.262 | eV |
| eps1 XY | 9.335 | 3.441 | |
| eps2 XY | 4.052 | 6.447 | |
| f2 XY | 1.774 | 0.046 | eV |
| s2 XY | 5.296 | 14.916 | eV$^2$ |
| d2 XY | 0.044 | 0.029 | eV |
| f3 XY | 2.038 | 0.071 | eV |
| s3 XY | 0.922 | 2.784 | eV$^2$ |
| d3 XY | 0.106 | 0.3 | eV |
| | | | |
| RMSE | 19.353 | | |

*Table 6: Fit parameters for anisotropic approach of 4 layer of MoS2.*

# Appendix E: Spectroscopic ellipsometry of few-Layer of MoS$_2$

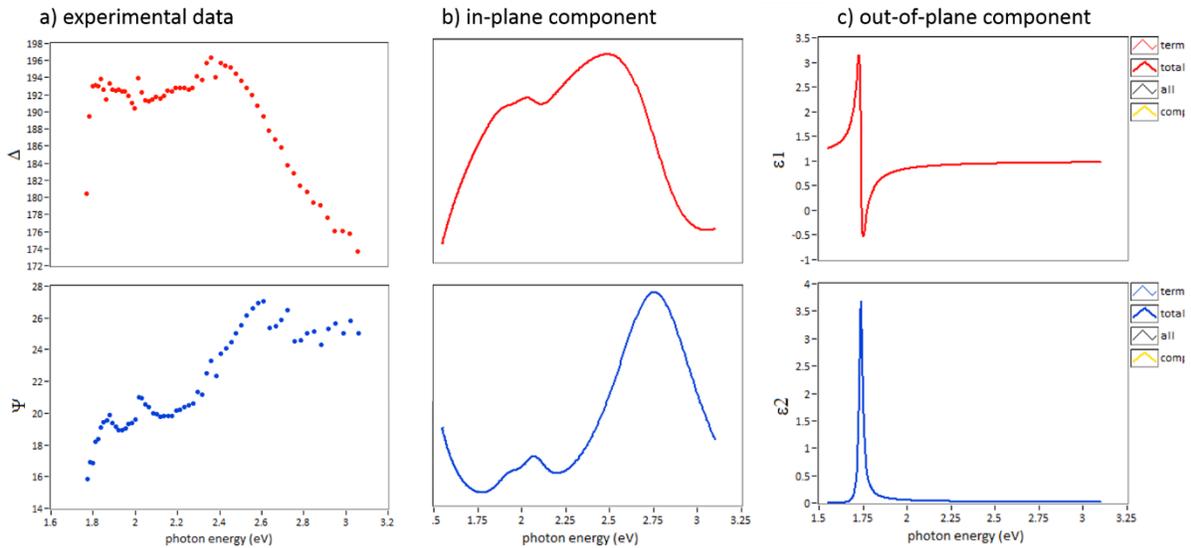

*Figure 5: (a) Measured ellipsometric angles for few layer MoS2 on sapphire (area on the flake is shown in figure 9). From a fit to the data using the anisotropic fit approach the in-plane component (b) and the out-of-plane component (c) of the dielectric function for few-layer MoS$_2$ are extracted.*

In Figure 5(a) Δ and Ψ values extracted from a few layer part of the flake shown in figure 9 and also in figure 3 of the manuscript are plotted. We can't exclude that there are some fluctuations in the number of layers in the analyzed few-layer part. The real and imaginary part of the dielectric tensor for few-layers of MoS$_2$ extracted from the anisotropic fit approach are shown in figure 5 (b) and (c) for both the in-plane and the out-of-plane components, respectively. The anisotropic fit approach results in a RMS value of 14.086, whereas the isotropic model approach (not shown) is significantly larger and constitutes 32.739.

All expected excitonic transitions can be seen in the extracted optical dispersions from both approaches. The ellipsometric angles and the in-plane as well as out of plane components of the dielectric tensor for mono-, tri- few layer and also the 4L part for an energy range between 2.1 eV and 3.2 eV are contrasted in appendix F.

# Appendix F: Comparative summary of the spectroscopic ellipsometry results

The measured ellipsometric angles as a function of photon energies for mono-, tri- few layer, bulk and also the 4L part are contrasted in figure 6. The extracted in-plane and out-of-plane components of the complex dielectric function $\varepsilon_{1,2}(x,y)$ and $\varepsilon_{1,2}(z)$, respectively are compared in figure 7. We note that the parasitic reflection from the side facet of the step from the small 4L region to the bulk part significantly impacts the measured values for the ellipsometric angles particularly in the spectral range of smaller photon energies. The values for those photon energies around the A and B excitons are excluded in the 4L fit to extract the dielectric function from the anisotropic fit to the data. Still, the amplitude of the extracted in-plane and out-of plane components of the complex dielectric function are not precise. A fact that is also indicated by the large RMSE of the fit model. The energetic position of the critical points in the extracted dispersion of the 4L part, however, seems to be reasonable, at least for the in-plane component of the dielectric tensor.

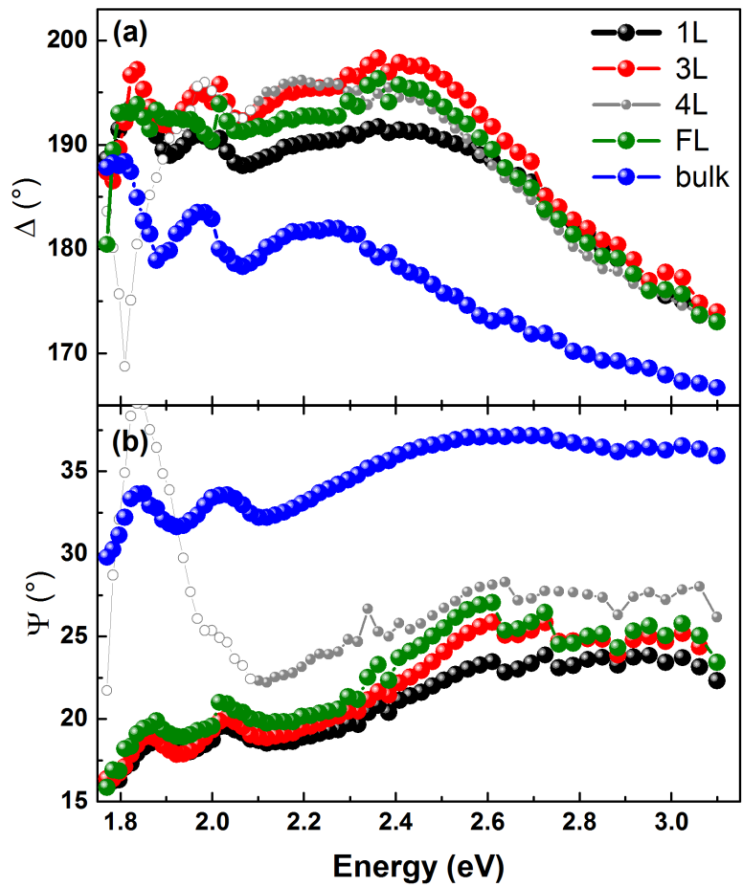

*Figure 6: Ellipsometric angles for mono-, tri-, four-, few-layer and bulk MoS$_2$ on sapphire of the flake displayed in figure 8 and in figure 3 of the manuscript as a function of the photon energy. The reflected signal of the four-layer part is strongly affected by parasitic side reflection from the large step-height from four-layer to bulk region of the flake in very close vicinity of the measured area of less than 2µm x 2µm in size that was in the limit of the resolution of the setup (<2µm). The impact of the spurious reflection from the side facet of the bulk step seems to be worse for larger photon energies (open grey circles) compared to smaller photon energies (filled grey spheres).*

The quantitative comparison of the dielectric functions for mono-, tri- and few-layer MoS$_2$ discloses the following trends:

- The energies of the A and B excitonic transitions observable in the in-plane components transition are rather independent by the number of layers, however the excitonic signature in the dielectric function, particularly at the A and B excitons, are weakened and broadened with increasing number of layers.
- The energies of the C-exciton transitions observable in the in-plane components are redshifted with increasing number of layers
- Overall, the amplitude of the in-plane components of the dielectric functions are reduced by increasing the number of layers indicating reduced light matter interaction.
- The critical points observable in the out-of-plane components of the dielectric function is redshifted with increasing the number of layers.
- The amplitude of the critical point increases with increasing number of layers indicating an increased light-matter interaction also perpendicular to the MoS2 layers consistent with increased thickness of the material in z-direction.

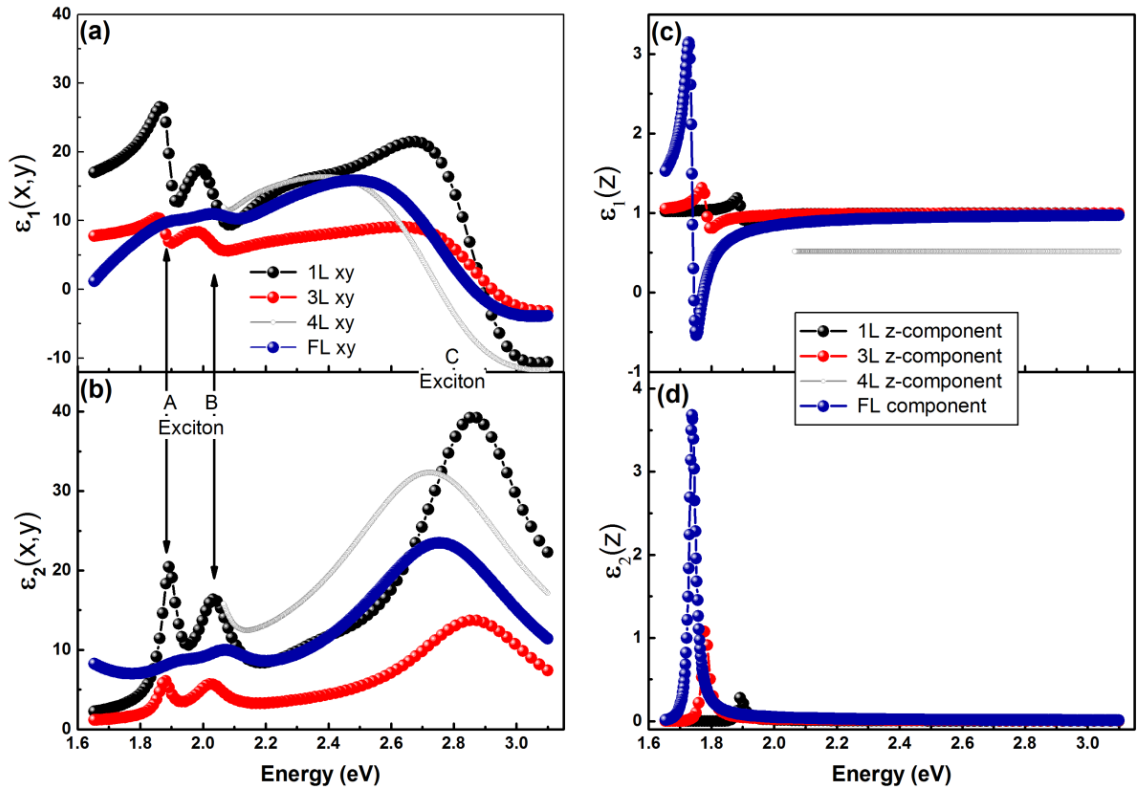

*Figure 7: Complex dielectric function of $MoS_2$ extracted from fits to the ellipsometric data in Fig.6. Real part of the in-plane component of the dielectric tensor $\varepsilon_1(x,y)$ from an anisotropic model for mono-, tri-, four- and few-layer $MoS_2$. (b) Imaginary part of the in-plane component of the dielectric tensor $\varepsilon_2(x,y)$ as in (a) and the data from an isotropic model for comparison. (c) Real part $\varepsilon_1(z)$ and (d) imaginary part $\varepsilon_2(z)$ of the dielectric tensor in out-of plane direction from the anisotropic model to the ellipsometry measurements.*

## Appendix G: Raman spectra and fit to Raman data of $MoS_2$ on $Si/SiO_2$

Raman measurements of the first order phonon modes, particularly the in-plane $E^1_{2g}$ phonon mode the LO phonon of the lattice and the homopolar out-of-plane $A_{1g}$ phonon mode provide access to a variety to parameters such as number of layers [2], doping density [3,4], temperature variation [5], disorder and lattice defects in the crystal [6] as well as strain [7, 8]. Consequently, Raman investigations of MoS$_2$ flakes is a very versatile, fast and non-destructive tool to characterize and analyze the crystal quality of MoS$_2$ flakes. Here, we briefly summarize, in which ways the two above described MoS$_2$ phonon modes will be affected by changing the number of layers, doping density, temperature, disorder, lattice defects, strain and also how the relative intensities of the phonon modes change with the excitation wavelength used for the Raman measurements.

- **Number of layers**: With decreasing number of layers the in-plane phonon mode $E^1_{2g}$ stiffens and is blue-shifted, whereas simultaneously the out-of-plane $A_{1g}$ mode weakens and is redshifted. As a result, the energy difference between the two Raman active modes $\Delta E=|E(A_{1g})-E(E^1_{2g})|$ is an excellent measure for the number of layers[2].

- **Charge carrier density:** With increasing electron density the $A_{1g}$ mode is redshifted and broadened due to phonon renormalization [3]. The in-plane $E^1_{2g}$ phonon mode is unaffected by variations in the doping density. The effect of doping on the $A_{1g}$ phonon mode can be found not only for monolayers, but also for bi- tri- and multilayers [4].

- **Temperature:** With increasing temperature, both first order phonon modes, $E^1_{2g}$ phonon mode and the $A_{1g}$ phonon are simultaneously redshifted due to the expansion of the lattice constant [5].

- **Disorder and lattice defects:** With increasing disorder and increasing density of lattice defects, the in-plane $E^1_{2g}$ phonon mode is redshifted and broadened, whereas the out-of-plane $A_{1g}$ mode is simultaneously blue-shifted and also broadened [6].

- **Strain:** Strain affects predominantly the in-plane $E^1_{2g}$ phonon mode that is shifted dependent on the applied compressive or tensile strain and can even be split into two modes in the presence of strain. Strain impacts the $A_{1g}$ phonon mode much less compared to the $E^1_{2g}$ phonon mode [7-9].

- **Mode intensity in dependence of excitation energy:** The mode intensities of both modes, the $E^1_{2g}$ and $A_{1g}$ phonon modes are not a measure for the number of layers, because already the peak as well as integrated intensities of monolayers can equal or even exceed those of thicker bulk regions (see e.g. [2] and figure 8). Whereas there is a clear trend from mono- to trilayers that the intensities increase for increasing number of layers. This trend does not hold by further increasing the number of layers. The relative mode intensities are strongly dependent of the excitations wavelength due to different resonance behavior for the in-plane and out-of-plane modes. The $A_{1g}$ mode is resonantly enhanced for excitation energy close to the A and B exciton, whereas the $E^1_{2g}$ is resonantly enhanced for excitation energy close to the C excitonic transition, caused by the symmetry of exciton-phonon interaction [10].

For the above given reasons, the investigation of the described phonon modes provide unique information not only to determine the number of layers, but even further to learn about the crystal quality and disorder, potential strain induced by the preparation of the flake by micromechanical cleavage, the intrinsic charge carrier density and their lateral variation. The Raman measurements performed in the presented work has been done with a high lateral resolution better than 1µm and is therefore also sensitive to lateral variations of the described parameters. In addition to the Raman measurements provided in the main manuscript, individual spectra, fits to the spectra using two Lorentzians and the fit parameters are summarized below for MoS$_2$ transferred on top of Si/SiO$_2$ substrate (figure 8 and table 7) and MoS$_2$ on sapphire substrate (figures 9, 10 and table 8). In comparison, one can see from the energy difference $\Delta E=|E(A_{1g})-E(E^1_{2g})|$ that the individual MoS$_2$ flake on Si/SiO$_2$ exhibits a higher intrinsic electron density compared to the MoS$_2$ flake in sapphire. Both flakes seem to be deposited without significant amount of strain despite wrinkled regions by the used micromechanical cleavage and transfer methods. In addition no signature for pronounced disorder and lattice defects can be found from the performed Raman investigations for the two investigated MoS$_2$ flakes.

# Appendix H: Raman spectra and fit to Raman data of MoS$_2$ on Si/SiO$_2$

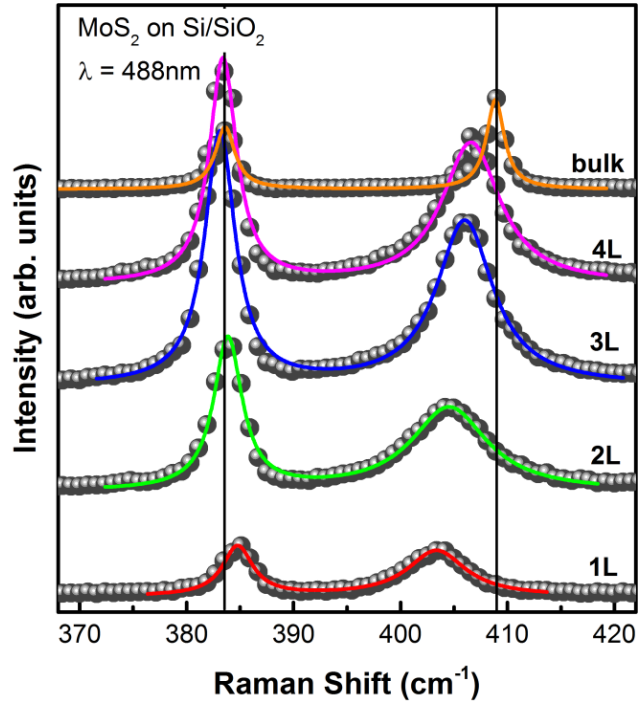

*Figure 8: Raman spectra (filled spheres) and fit to the spectra using two Lorentzians (solid lines) for MoS$_2$ on Si/SiO$_2$ substrate. The Flake is displayed in Fig. 2(a) of the manuscript. The spectra from the upper monolayer and bilayer parts are displayed. Fits parameters are given below in the table also from the lower mono- and bilayer region on the flake.*

| MoS$_2$ on Si/SiO$_2$: Fit Parameter from Raman measurements | | | | | | | |
|---|---|---|---|---|---|---|---|
| | E$^1_{2g}$-phonon mode | | | A$_{1g}$ phonon mode | | | |
| # of layer | *Mode energy (cm$^{-1}$)* | FWHM (cm$^{-1}$) | Int. Intensity (arb units) | *Mode energy (cm$^{-1}$)* | FWHM (cm$^{-1}$) | Int. Intensity (arb units) | ΔE (cm$^{-1}$) |
| 1L upper | 384.8 (+/-0.05) | 4.43 (+/-0.17) | 3.1 (+/-0.14) | 403.4 (+/-0.07) | 6.66 (+/-0.006) | 5.5 (+/-0.26) | 18.6 |
| 1L lower | 384.4 (+/-0.04) | 3.26 (+/- 0.14) | 3.8 (+/-0.14) | 403.1 (+/-0.05) | 5.78 (+/- 0.007) | 6.7 (+/-0.23) | 18.3 |
| 2L upper | 383.8 (+/-0.03) | 3.57 (+/-0.12) | 10.22 (+/-0.3) | 404.6 (+/-0.09) | 8.36 (+/-0.37) | 14.0 (+/-0.65) | 20.8 |
| 2L lower | 383.9 (+/-0.03) | 3.22 (+/-0.10) | 8.9 (+/- 0.2) | 404.6 (+/- 0.09) | 8.66 (+/-0.36) | 13.0 (+/-0.6) | 20.7 |
| 3L | 383.1 (+/-0.04) | 3.57 (+/-0.1) | 15.9 (+/-0.4) | 406.0 (+/-0.06) | 6.44 (+/-0.23) | 19.3 (+/-0.65) | 22.9 |
| 4L | 383.1 (+/-0.04) | 3.11 (+/-0.1) | 12.7 (+/-0.4) | 406.6 (+/- 0.08) | 6.61 (+/-0.3) | 17.0 (+/-0.70) | 23.5 |
| bulk | 383.5 (+/-0.04) | 2.24 (+/-0.1) | 2.5 (+/-0.1) | 408.9 (0.02) | 2.1 (+/-0.1) | 0.33 (+/-0.1) | 25.4 |

*Table 7: Fit parameter from the Lorentzian fit to the Raman spectra of the MoS$_2$ flake on Si/SiO$_2$ substrate displayed in the manuscript figure 2(a).*

# Appendix I: Raman spectra and fit to Raman data of MoS$_2$ on sapphire

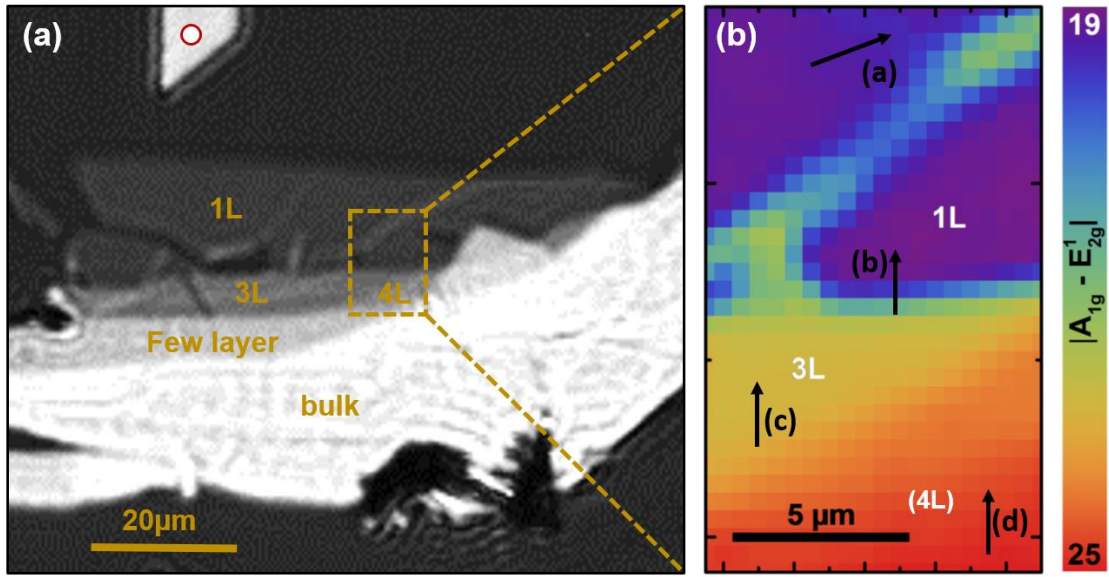

*Figure 9: (a) IE contrast image of MoS$_2$ flake on sapphire with mono-, tri-, four-, few-layer parts and a bulk region. The spectroscopic ellipsometry measurements have been done on this flake. (b) Raman map with the energy difference $\Delta E=|E(A_{1g})-E(E^1_{2g})|$ between the mode energies of the $A_{1g}$ and the $E^1_{2g}$ phonon mode from fits to the data with two Lorentzian color coded. The arrows mark the position of which the individual spectra are taken that are displayed in figure 10.*

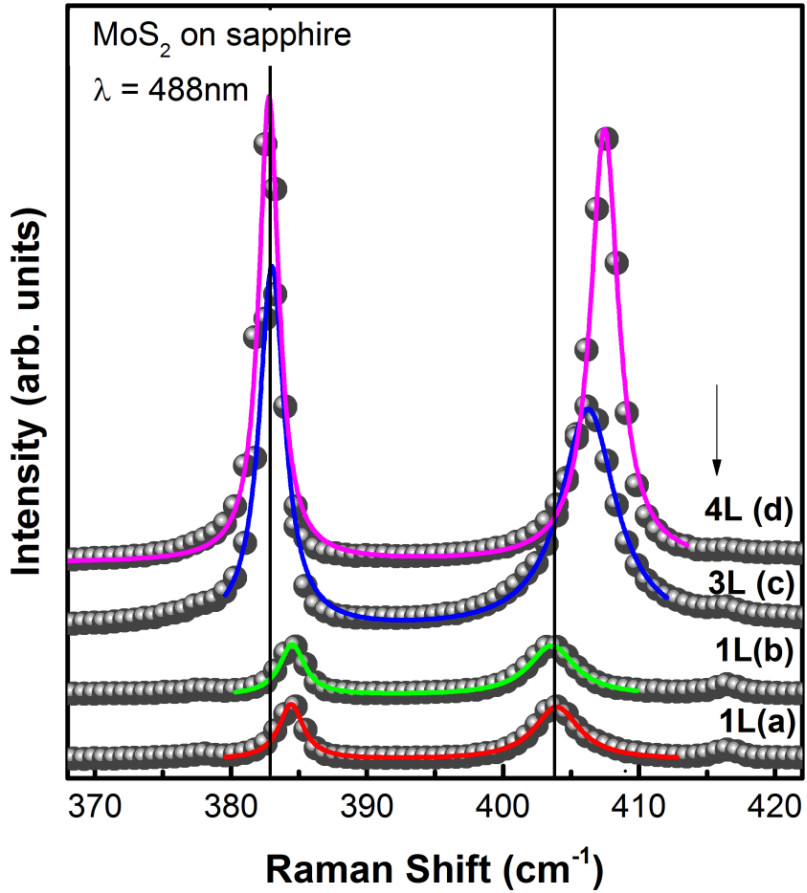

Figure 10: Raman spectra (filled spheres) and fit to the spectra using two Lorentzians (solid lines) for MoS$_2$ on sapphire substrate. An image of the flake is displayed in figure 9 and figure. 3 of the manuscript. The arrow marks a Raman signal from the sapphire substrate.

| MoS$_2$ on saphhire: Fit Parameter from Raman measurements | | | | | | | |
|---|---|---|---|---|---|---|---|
| | $E^1_{2g}$-phonon mode | | | $A_{1g}$ phonon mode | | | |
| # of layer | Mode energy (cm$^{-1}$) | FWHM (cm$^{-1}$) | Int. Intensity (arb units) | Mode energy (cm$^{-1}$) | FWHM (cm$^{-1}$) | Int. Intensity (arb units) | ΔE (cm$^{-1}$) |
| 1L (a) | 384.4 (+/-0.03) | 2.20 (+/-0.11) | 1.3 (+/-0.05) | 404.0 (+/-0.04) | 4.07 (+/-0.01) | 2.3 (+/-0.09) | 19.6 |
| 1L (b) | 384.5 (+/-0.02) | 2.25 (+/- 0.08) | 1.2 (+/-0.04) | 403.6 (+/-0.03) | 4.2 (+/- 0.01) | 2.3 (+/-0.06) | 19.1 |
| 3L (c) | 383.0 (+/-0.02) | 2.2 (+/-0.06) | 8.7 (+/-0.2) | 406.2 (+/-0.04) | 4.8 (+/-0.16) | 11.5 (+/-0.35) | 23.2 |
| 4L (d) | 382.8 (+/-0.02) | 1.8 (+/-0.06) | 9.0 (+/-0.2) | 407.5 (+/-0.02) | 2.5 (+/-0.06) | 11.8 (+/-0.2) | 24.7 |

Table 8: Fit parameter from the Lorentzian fit to the Raman spectra displayed in figure 10 on the MoS$_2$ flake on sapphire displayed in figure 9 and manuscript figure 2(a).